\documentclass[doublespacing]{elsart}
\usepackage[T1]{fontenc}
\usepackage{graphicx}
\usepackage{amssymb}
\usepackage{amsmath}
\usepackage{cite}
\providecommand{\tabularnewline}{\\}

\begin{document}
\begin{frontmatter}
\title{Identification of Photon-tagged Jets in the ALICE Experiment}

\author[a,b]{G. Conesa\corauthref{cor1}} 
\corauth[cor1]{Corresponding author}
\ead{Gustavo.Conesa.Balbastre@cern.ch} 
\author[b]{ H. Delagrange}
\author[a]{J. D\'{\i}az}
\author[d]{ Y.V. Kharlov}
\author[b,e]{Y. Schutz}
\address[a]{IFIC (Centro Mixto Universidad de Valencia-CSIC), Valencia, Spain}
\address[b]{SUBATECH, Ecole des Mines, IN2P3/CNRS, Universit\'e de Nantes, Nantes, France}
\address[d]{Institute for High-Energy Physics, Protvino, Russia}
\address[e]{CERN, Gen\`eve, Switzerland}

\begin{abstract}
{The ALICE experiment at LHC will detect and identify
prompt photons and light neutral-mesons with the PHOS detector and the additional
 EMCal electromagnetic calorimeter. Charged particles will
be detected and identified by the central tracking system. In this article, the possibility of studying the interaction of  jets with the nuclear medium, using prompt photons as a tool 
to tag jets, is investigated by simulations. New methods to identify prompt photon-jet
events and to distinguish them from the jet-jet background are presented.}
\end{abstract}
\end{frontmatter}

%-----------------------------------------------------------

\section{Introduction\label{intro}}
In Ref.~\cite{ICM}, we stressed the importance of the measurement of 
jets in the ALICE experiment at  LHC to study the properties of 
the nuclear medium predicted to be created in ultra-relativistic 
heavy-ion collisions, the Quark-Gluon Plasma
\cite{Accardi:2003gp,Bjorken:1982tu}. Jets will be abundantly 
produced at the LHC ($2 \times 10^{6}$ jets with $p_{T}>100$~GeV/$c$ per 
year in the ALICE acceptance) enabling inclusive and exclusive jet measurements. In particular, 
jet topology (jet shape, jet \emph{heating}, fragmentation function,\ldots{}) 
can be measured to study the redistribution of the jet  energy among the 
fragmentation particles after the jet has traversed the nuclear medium 
created in the collision~\cite{Salgado:2003rv}. These studies require
the identification of jets and the measurement, as accurately as possible,
of the jet energy, ideally before and after interaction with the medium. 
A very attractive method to perform these studies is to tag jets with
prompt photons emitted  opposite  to the jet direction.
The dominant processes for such events are $g+q\rightarrow\gamma+q$
(Compton) and $q+\bar{q}\rightarrow\gamma+g$ (annihilation), although
recent studies show that Next to Leading Order processes (NLO) contribute 
significantly to the photon spectrum  below 50~GeV/$c$ \cite{Arleo:GTagged}. As photons
emerge almost unaltered from the dense medium, they  provide a measurement
of the original energy of the parton emitted in the opposite direction.
On one hand, this coincidence technique  can be used  to localize the
jet and on the other hand  it allows to build the parton fragmentation
function without the need of reconstructing the jet energy from the detected
hadrons. Medium effects can then be identified through modifications
of the fragmentation function, i.e., by the redistribution of the jet
energy among its components.

The identification of prompt photons in
ALICE with the high resolution photon spectrometer PHOS in association with
the central tracking system has been presented earlier~\cite{ICM,PPRv2}. 
The identification of prompt-photon jet events in ALICE  is found optimal for photons 
with energy larger than 20~GeV. Below this energy, decay and prompt photons cannot 
be efficiently separated on an event by event basis. 

In this paper, we discuss the possibilities offered by the addition of the EMCal electromagnetic 
calorimeter to the ALICE setup. We present an algorithm for identifying photon-jet 
events and for reconstructing hadron jet features. The algorithm was developed and tested with 
the simulations described in Ref.~\cite{ICM}.  The study was done 
with generated particles reconstructed with a fast reconstruction algorithm, due to lack of computer 
time for a full reconstruction. The generator used to produce  $pp$ collisions 
at $\sqrt{s}=5.5$~TeV is PYTHIA  6.203~\cite{Sjostrand:2000wi,Sjostrand:2001yu}. The simulations 
generated with PYTHIA contain $\gamma$-jet and jet-jet events, with prompt and decay photons, 
respectively, in the energy range 20<E<100~GeV.  The generator used to reproduce the underlying event
of a heavy-ion collision is HIJING 1.36~\cite{Gyulassy:1994ew}. 
To show the improvements provided by 
the electromagnetic calorimeter EMCal in the jet reconstruction procedure, two possible 
experimental situations are considered: i) only charged particles can be detected (EMCal 
is not present) and ii) neutral particles can also be detected (EMCal is present).

\section{Acceptance and response of the detectors\label{GAMMAJET:FastResponse} }

To assess the performance of our photon-jet identification algorithm, 
a fast detector simulation is used because a complete event simulation with 
particle transport and reconstruction would have required an unaffordable 
computing time. In this fast simulation framework, we rely only on the
properties of the final state particles generated by PYTHIA and on
the knowledge of the response function and acceptance of the various detectors involved.

\begin{figure}
\begin{center}\includegraphics[%
  width=6cm,
  keepaspectratio]{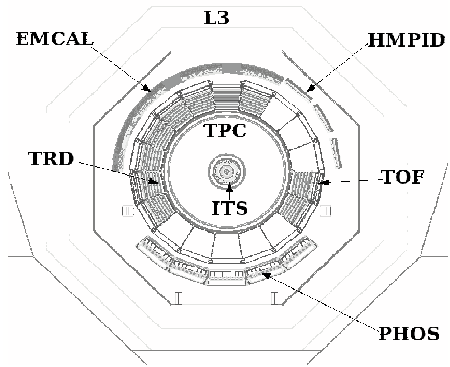}\end{center}
\caption{\label{alicelayout} Sketch of the ALICE experiment, transverse view.}
\end{figure}

The ALICE experiment is sketched in  Fig.~\ref{alicelayout}. The acceptances of all the 
detectors relevant  for  this study are listed in Table~\ref{cap:TabAcceptance}. A full 
description of the ALICE detector can be found in Ref.~\cite{PPRv1}.

The response function of the detectors involved in our analysis,
PHOS and the central tracking system (TPC), are described in Sec.~3 of 
Ref.~\cite{ICM}. The EMCal response was assumed to be identical to the PHOS
response, although  in fact it will be slightly worse.
However, since the EMCAL granularity is coarser, 
the $p_T$ range over which a Shower Shape Analysis (SSA) can discriminate  prompt photons 
from decay photons\footnote{Due to Lorentz contraction and the calorimeters 
granularity decay photons merge in one detected cluster for $E_{\pi^{0}} \gtrsim 30$ GeV in PHOS 
and  $E_{\pi^{0}} \gtrsim 10$ GeV in EMCal.}%
 is reduced from about $100$~GeV$/c$ in PHOS down to $40$~GeV$/c$ in EMCal.

\begin{table}

\caption{{\small \label{cap:TabAcceptance} TPC (Time Projection Chamber), PHOS and 
EMCal acceptances. The physical TPC $\eta$ acceptance is larger $(|\eta|<0.9)$, but
we take this lower value to ensure a good track matching. The acceptance of EMCal used 
in this study might not be the final one as the layout is still under discussion.}}

\begin{center}\begin{tabular}{cccc}

Detector~~&
$\left|\eta\right|$~~~~&
$\phi_{min}$~~~~&
$\phi_{max}$\tabularnewline
\hline
\hline 
PHOS&
0.12&
$220^\circ$&
$320^\circ$\tabularnewline
EMCAL&
0.7&
$60^\circ$&
$180^\circ$\tabularnewline
TPC&
0.7&
$0^\circ$&
$360^\circ$\tabularnewline
\hline
\end{tabular}\end{center}

\end{table}

\section{Jet selection procedure\label{GAMMAJET:selection}}

Photon-jet events were identified by applying the following $\gamma$-tagging
algorithm:

\begin{enumerate}
\item Find the most energetic prompt photon identified
in PHOS (Ref.~\cite{ICM}). 
\item Find the jet leading particle%
\footnote{Jets always have a  particle  carrying a significant fraction
of the jet energy (in average 40\%).}, either a charged hadron detected by the central tracking system or a neutral pion detected 
by the EMCal, with the highest $p_{T}$ value and emitted back-to-back to 
the photon ($\Delta\phi \sim 180^\circ$).
Neutral pions are identified from their two photon decay from all photons 
in the event with a relative angle between the limits defined by the decay kinematics
and with an invariant mass in the range $120<M_{\gamma\gamma}<150$~MeV/$c^{2}$.
In addition, it is required that the $p_T$ value of the leading particle be
at least 10\% of the photon energy. 
\item Reconstruct the jet as the ensemble of all particles above a given $p_T$ threshold 
contained inside a cone of a given radius around the leading particle direction. 

\item Finally, the event is identified as  a photon-jet if the ratio
of the reconstructed jet energy and the prompt photon energy falls
within a given selection window. 
\end{enumerate}

There are standard  algorithms that can be used to reconstruct  jets 
($k_{T}$, cone, etc), but they fail to reconstruct jets with energy smaller than 
40-50 GeV in a heavy-ion environment~\cite{Cormier:JetEMCAL, PPRv2}. 
Studies about tagging 
reconstructed jets with prompt photons  with standard  algorithms are in progress. 

\subsection{ Leading particle selection\label{GAMMAJET:anglecorrel}}
In ALICE, since the central tracking system and EMCal do not cover the same acceptance, 
jets can be fully reconstructed (charged plus neutral hadrons) only over a limited solid angle
($\left|\eta\right|<0.7$ and $\Delta \phi = 120^{\circ}$ in this study\footnote{The EMCal design 
is still under discussion and now, an azimuthal aperture of $110^{\circ}$ is considered.}).
Therefore, outside the common acceptance region, the jet energy is only partially
reconstructed which may lead to the  rejection of a true $\gamma$-jet event.

\begin{figure}
\begin{center}\includegraphics[
  width=6cm,
  keepaspectratio]{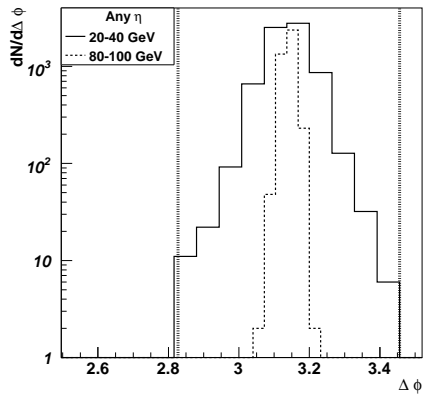}\end{center}

\begin{center}\includegraphics[%
  width=6cm,
  keepaspectratio]{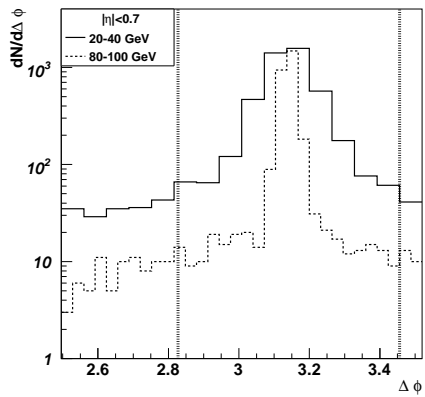}\end{center}

\caption{\label{GAMMAJET:phifig}{\small Azimuthal angle correlations between
prompt photons and their corresponding jet leading particles as a
function of $\Delta\phi=\phi_{l}-\phi_{\gamma}$ for simulated $\gamma$-jet
events in the energy ranges 20-40 GeV and 80-100 GeV for} \emph{\small pp}
{\small collisions. Jet particles (lower frame) are filtered through
the acceptance $\left|\eta\right|<0.7$. Dotted vertical lines represent the cut between  0.9$\pi$ 
and 1.1$\pi$ that we use to select  leading particles.}}
\end{figure}

The prompt photon and the parton at the origin of the observable $\gamma$-jet event 
are emitted in opposite directions in the center of mass of the hard process.
In the laboratory system,  the correlation in pseudo-rapidity is
washed out, as illustrated in Fig.~\ref{GAMMAJET:phifig}, while the azimuthal
angle correlation is conserved since there is no boost in the transverse direction.
The relative azimuthal angle, $\Delta\phi$, of $\gamma$-jet events is peaked at 
180${}^\circ$ and its width depends on both the detector acceptance and resolution, 
and  the energy of the event. We have tuned our algorithm to select photon 
leading-particle pairs satisfying the angular condition $0.9\,\pi<\Delta\phi<1.1\,\pi$.

\begin{figure}
\begin{center}\includegraphics[%
  width=6cm,
  keepaspectratio]{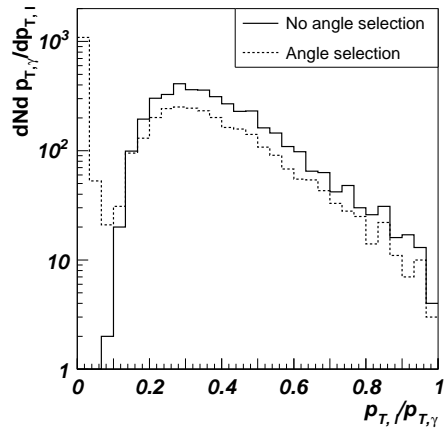}\end{center}

\caption{\label{GAMMAJET:leadratiofig}{\small Leading particle distribution
of $\gamma$-jet events as a function of the $p_{T,l}/E_{\gamma}$
ratio for $pp$ collisions and jet energies in
the range 80 -100 GeV. The dotted line corresponds to the case where the leading
particle is searched inside the acceptance of the detectors and opposite
in $\phi$ angle to the prompt photon. The solid line corresponds to the case 
whithout  acceptance restrictions.}}
\end{figure}

When the leading particle of a given event escapes from the detector
acceptance, the algorithm finds a wrong leading particle. This misidentification
produces the peak in the jet leading particle distribution at low values of the 
$p_{T,l}/E_{\gamma}$ ratio (Fig.~\ref{GAMMAJET:leadratiofig}),
where $p_{T,l}$ and $E_{\gamma}$ are the momentum of the identified
leading particle and the energy of the photon, respectively. These fake
leading particles are rejected by imposing the condition $p_{T,l}/E_{\gamma}>0.1$.
The probability of finding the jet associated to a prompt
photon is about 50\%, determined only by the jet detection acceptance.

\subsection{Identification of $\pi^{0}$ leading particles\label{GAMMAJET:pi0}}

\begin{figure}
\begin{center}\includegraphics[%
  width=6cm,
  keepaspectratio]{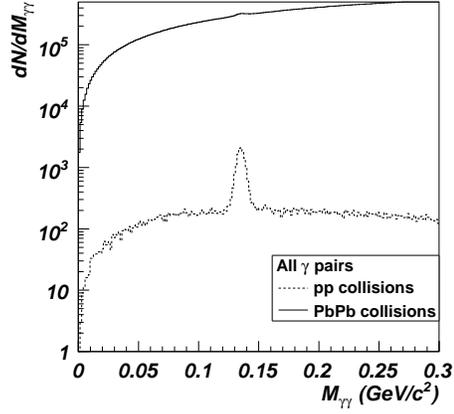}\end{center}

\caption{\label{GAMMAJET:noselectioncuts}{\small Invariant mass distributions
of all photon pairs in the event for $pp$ (dashed line) and Pb-Pb (solid line) collisions 
at $\sqrt{s}=5.5A$~TeV. No selection is applied.}}
\end{figure}

The $\pi^{0}$ candidates to leading particles are identified by detecting
their two decay photons in EMCal and selecting those with an invariant mass 
$M_{\gamma\gamma}$ around the $\pi^{0}$ rest mass. Because of the combinatorial background, 
which is particularly large in the case of Pb-Pb collisions (Fig.~\ref{GAMMAJET:noselectioncuts}), 
additional selection conditions are necessary.

\begin{figure}
\begin{center}\includegraphics[%
  width=6cm,
  keepaspectratio]{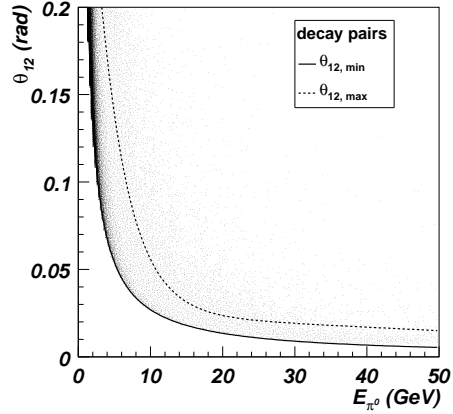}\end{center}

\caption{\label{GAMMAJET:anglecutfig}{\small Opening angle of $\pi^{0}$
decay photon pairs (without any restriction in detector acceptance). The lines
limit the opening angle range selected; the lower line is obtained
from the decay kinematics, the upper one has been empirically chosen
to select most of $\pi^{0}$ decay pairs.}}
\end{figure}
The opening angle  $\theta_{12}$ between the two  $\pi^{0}$ decay  photons is given
in the lab system by:

\begin{equation}
\cos\theta_{12}=\frac{\gamma_{\pi}^{2}\beta_{\pi}^{2}-\gamma_{\pi}^{2}\alpha^{2}-1}{\gamma_{\pi}^{2}(1-\alpha^{2})}\label{open},\end{equation}
where $\alpha$ is the decay asymmetry, $\gamma,$ the Lorentz factor
and $\beta$ the $\pi^{0}$ velocity in units of $c$. The opening angle is minimum
for symmetric decays ($\alpha=0$ ). From the simulated $\pi^{0}$ opening angle
distributions in Pb-Pb collisions (Fig.~\ref{GAMMAJET:anglecutfig}), the maximum opening 
angle was empirically selected by fitting  the equation
 \begin{equation}
\theta_{max}=0.4\cdot e^{-0.25\cdot E}+0.025-2\cdot10^{-4}\cdot E_{\pi^{o}}.\label{maxopen}\end{equation}
to the data.
\begin{figure}
\begin{center}\includegraphics[%
  width=6cm,
  keepaspectratio]{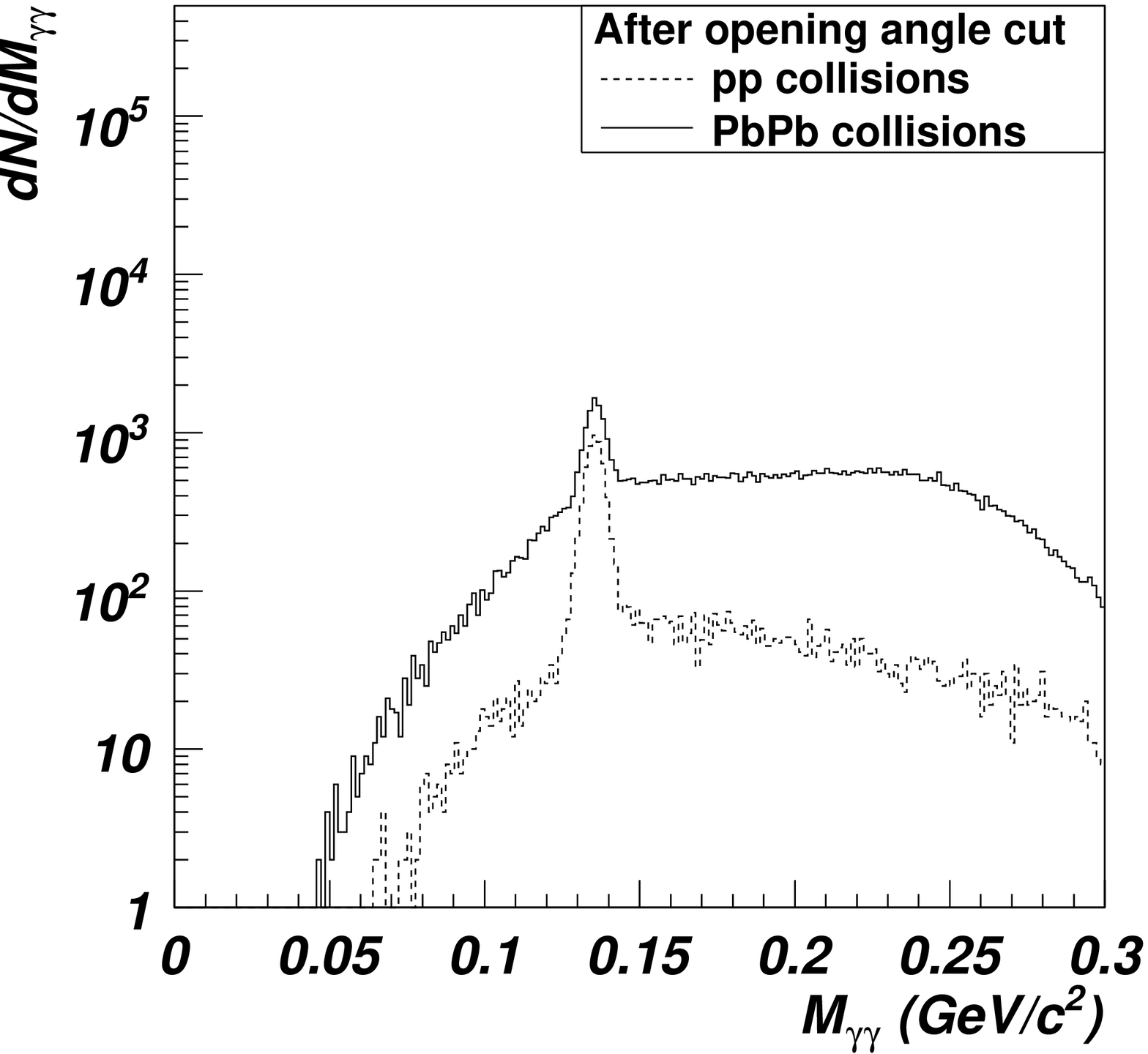}\end{center}

\begin{center}\includegraphics[%
  width=6cm,
  keepaspectratio]{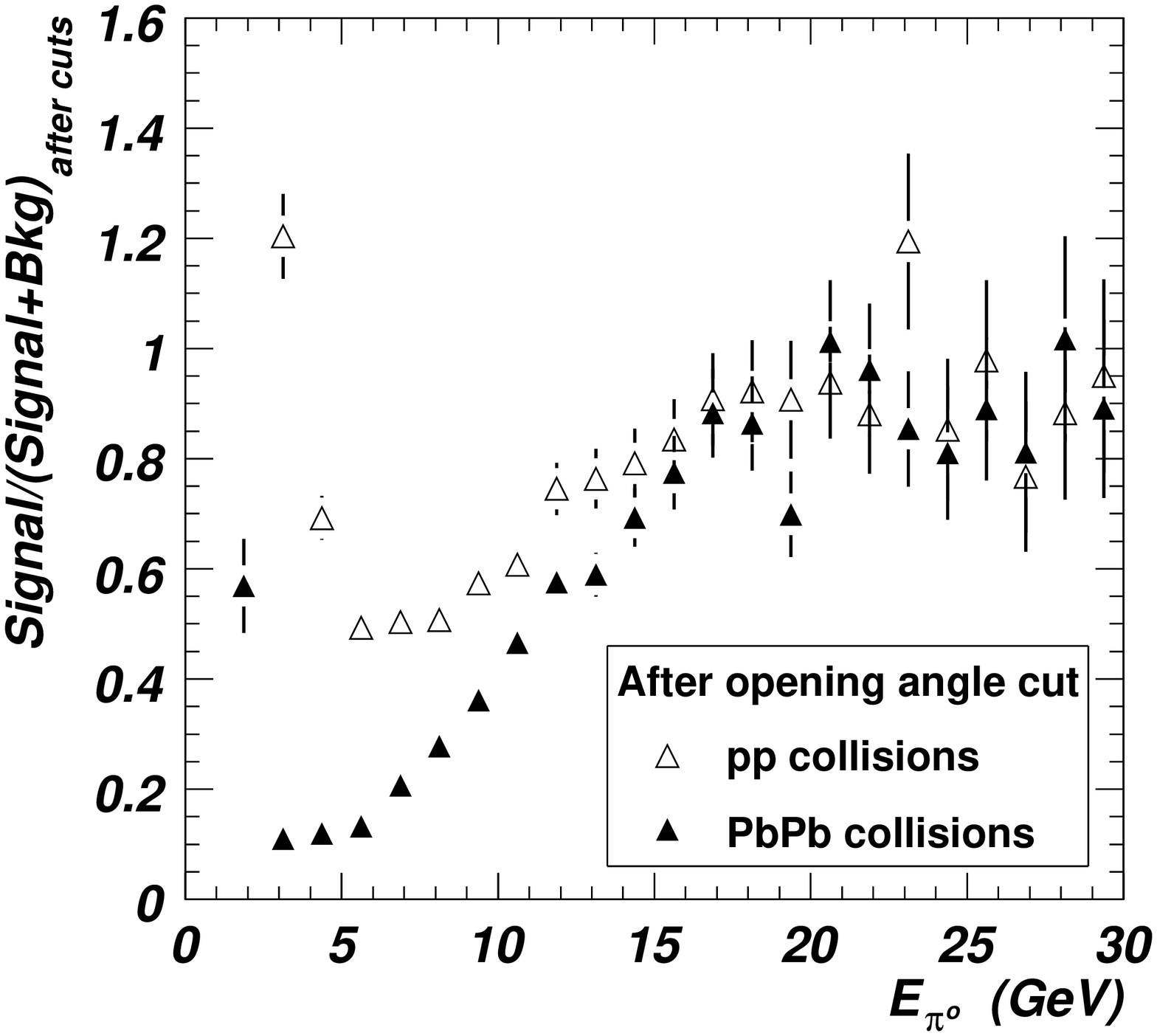}\end{center}

\caption{\label{GAMMAJET:selectioncutsfig}{\small Upper frame: Invariant
mass distributions of photon pairs in the event which satisfy the
$\phi$ and leading particle energy cuts and the opening angle restriction,
for $pp$ (dashed line) and Pb-Pb (solid line) collisions at  $\sqrt{s}=5.5A$~TeV. 
Lower frame: Ratio of the simulated $\pi^{0}$ number to the selected photon 
pair number in the event, for $pp$~($\vartriangle$) and Pb-Pb~($\blacktriangle$) 
collisions at $\sqrt{s}=5.5A$~TeV. Only $\gamma$-jet events with jet energy in the range from 
20 to 100~GeV are considered. }}
\end{figure}

\begin{figure}
\begin{center}\includegraphics[%
  width=6cm,
  keepaspectratio]{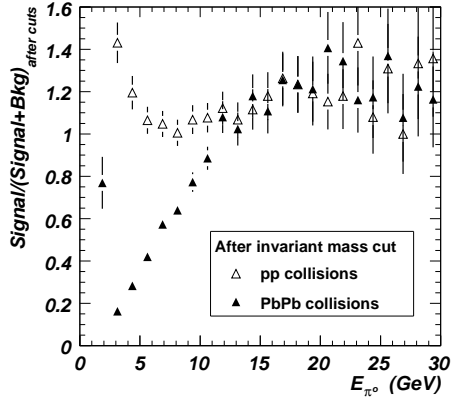}\end{center}

\caption{\label{GAMMAJET:pi0allcutsfig}{\small Ratio of number of simulated
$\pi^{0}$ to the total number of photon pairs satisfying the opening
angle, invariant mass and leading particle selection conditions for $pp$
 ($\vartriangle$) and Pb-Pb ($\blacktriangle$) collisions at $\sqrt{s}=5.5A$~TeV. 
Only $\gamma$-jet events in the jet energy range from 20 to 100~GeV are considered. }}
\end{figure}

The condition $\theta\le\theta_{max}$ selects more than 80\% of the $\pi^{0}$
decay photon pairs\footnote{The selection efficiency decreases from about 92\% at 
$E_{\pi}\sim2$~GeV to  around 82\%  at $E_{\pi}\sim15$~GeV and then increases again 
to values larger than 90\% for  $E_{\pi}>30$~GeV.} and rejects more than
the 93\% of all the uncorrelated photon pairs (Fig.~\ref{GAMMAJET:selectioncutsfig}). 
$\pi^0$ are finally identified by requiring the additional condition 
$120<M_{\gamma\gamma}<150$~MeV/$c^{2}$ on the invariant mass. The efficiency of the $\pi^0$ identification 
is calculated from the ratio of the number of simulated $\pi^{0}$ to the whole 
number of photon pairs produced in the heavy-ion event satisfying the selection 
conditions (Fig.~\ref{GAMMAJET:pi0allcutsfig}).
This ratio is between 1 and 1.2 for $\pi^{0}$ of energies greater
than 5~GeV for \emph{pp} collisions at $\sqrt{s}=5.5$~TeV and  10~GeV for
Pb-Pb collisions at $\sqrt{s}=5.5A$~TeV. The ratio may be larger than one because the opening
angle restriction eliminates a fraction of true $\pi^{0}$
decay  pairs.   The combinatorial background is responsible of  the fast decrease 
 of the signal to noise ratio in Pb-Pb collisions at low energies
(Fig.~\ref{GAMMAJET:pi0allcutsfig}).

\section{Jet reconstruction\label{GAMMAJET:jetrec} }

Jets are reconstructed starting from the seed provided by the leading
particle found as described above. Particles found in a cone around the leading particle
at ($\eta_{l},\:\phi_{l}$), of size $R$ defined by the equation,

\begin{equation}
R=\sqrt{(\phi_{l}-\phi)^{2}+(\eta_{l}-\eta)^{2}},\label{cone}\end{equation}
are assigned to a jet if their transverse momentum exceeds a given $p_T^{th}$. 
Again, a fraction of jet particles may be lost due to the
limited detector acceptance. Only jets with reconstructed energy
comparable to the energy of their corresponding prompt photon are
selected.

\begin{figure}
\begin{center}\includegraphics[%
  width=5.5cm,
  keepaspectratio]{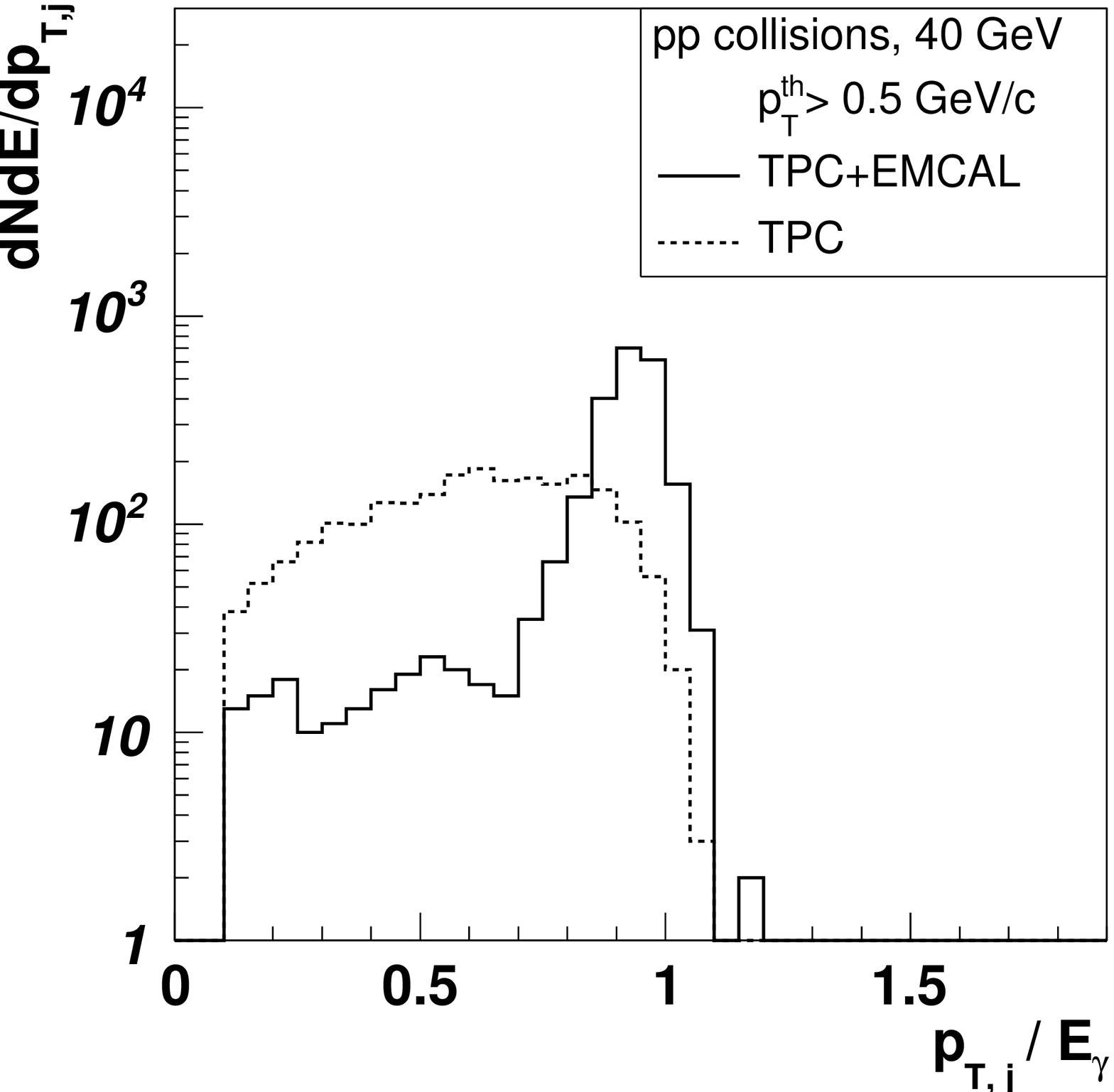}\end{center}

\begin{center}\includegraphics[%
  width=5.5cm,
  keepaspectratio]{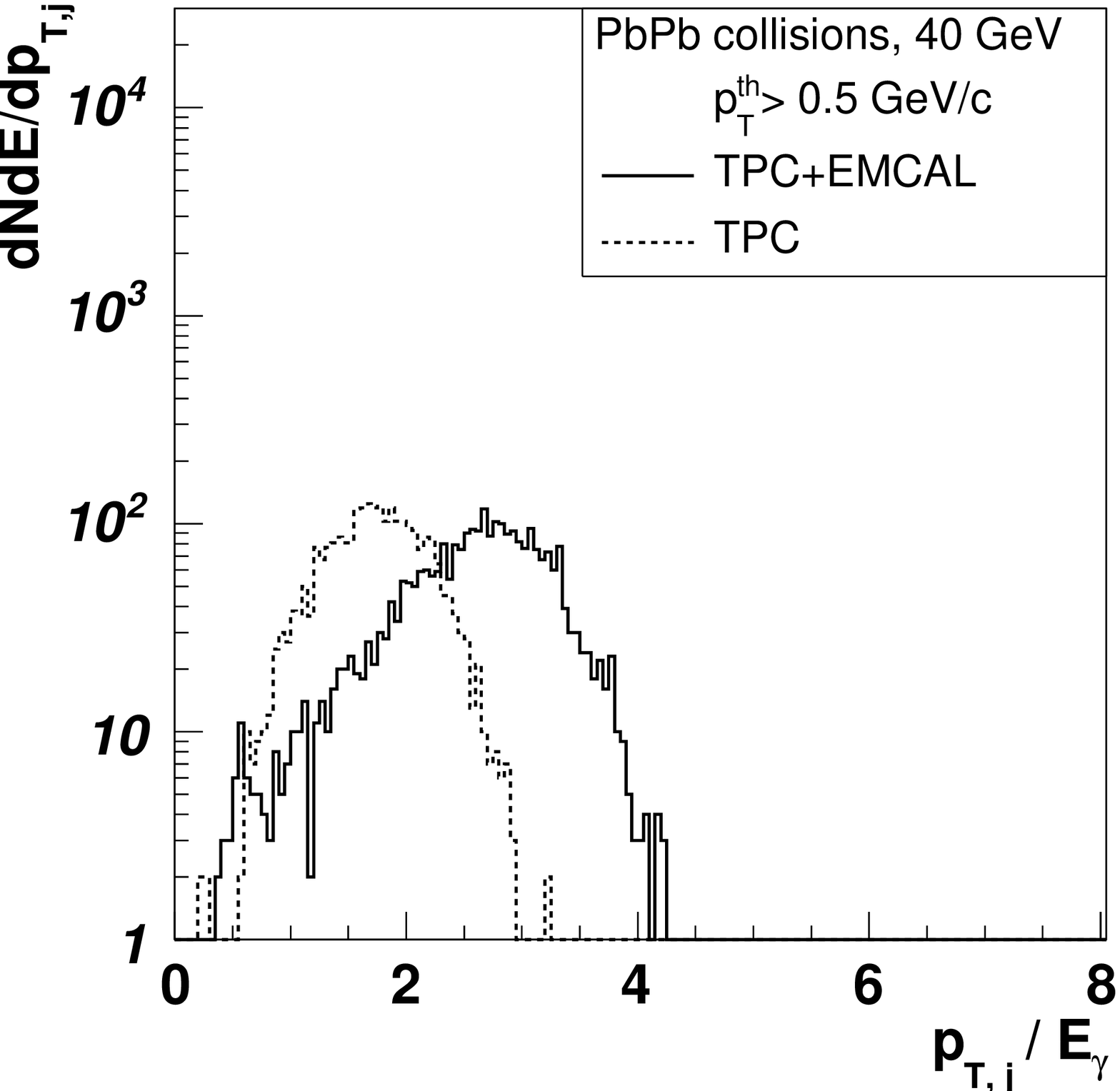}\end{center}

\begin{center}\includegraphics[%
  width=5.5cm,
  keepaspectratio]{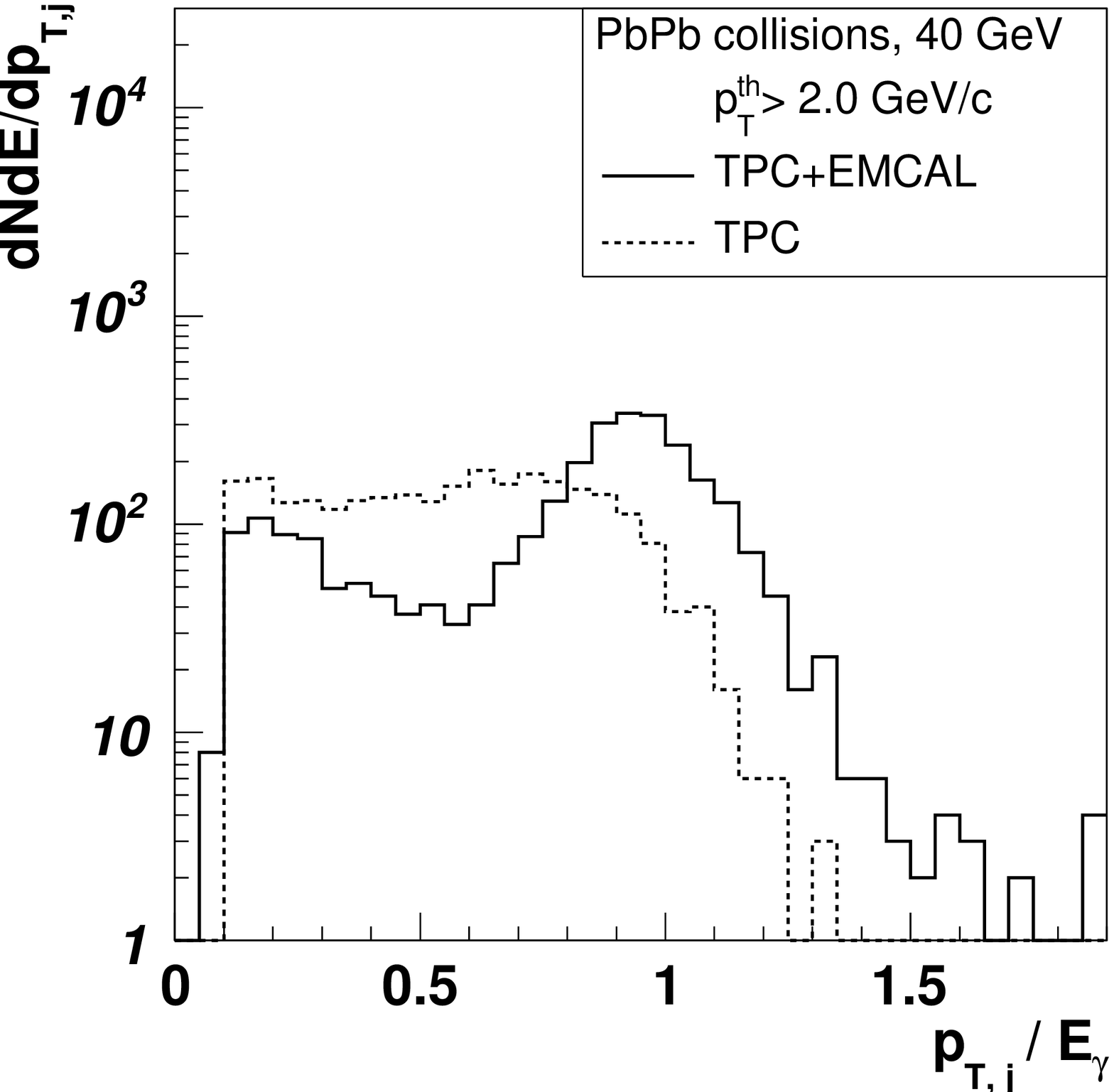}\end{center}

\caption{\label{GAMMAJET:jetratio}{\small Jet distribution as a function
of the ratio $p_{T,j}/E_{\gamma}$ for generated $\gamma$-jet events 
in $pp$ collisions (top frame) and Pb-Pb collisions (middle and bottom frames) 
at $\sqrt{s}=5.5A$~TeV.  A jet cone of $R=0.3$ and a jet particle threshold of 
$p_{T}^{th}=0.5$~GeV/$c$ are required in the top and middle frames and a jet particle 
threshold of $p_{T}^{th}=2$~GeV/$c$ is taken in the lower one. Results for the setups 
without EMCal (dashed line) and with EMCal (solid line) are given.}}
\end{figure}

In our jet finding algorithm, two experimental configurations are
considered,

\begin{enumerate}
\item Charged particles are detected in the central tracking system and
neutral particles in EMCal. This configuration is labeled TPC+EMCal
in figures and text.
\item Only the central tracking system is available and consequently only
charged particles can be detected; this configuration is labeled
TPC in figures and text.
\end{enumerate}
We compare in the upper part of Fig.~\ref{GAMMAJET:jetratio} the ratio of the
measured jet transverse momentum $p_{T,j}$  ($p_T^{th}$=0.5~GeV/$c$, $R=0.3$) 
to the energy of the corresponding prompt photon $E_{\gamma}$ for \emph{pp} collisions 
and  $E_{\gamma}=40$~GeV and for both detector configurations. 
In the TPC+EMCal configuration, the jet energy is correctly reconstructed, and values 
close to the energy of the corresponding prompt photon are obtained. 
As  the energy carried away by neutral particles can only be detected with  EMCal, 
the $p_{T,j}/E_{\gamma}$ distribution does not show the expected correlation 
in the TPC configuration. The correlation is best observed for the highest jet energy 
studied. This result is independent of the cone size.

In the case of Pb-Pb collisions, the background is large
and the $p_{T,j}/E_{\gamma}$ distributions are wide and peak at values
greater than unity (middle part of Fig.~\ref{GAMMAJET:jetratio}).
Requiring a higher $p_{T}$ threshold than for \emph{pp} collisions ($p_{T}^{th}=2$~GeV/$c$)
to reduce the background, the distribution features (peak position and width) resemble
those obtained for $pp$ collisions, at least for high-energy jets 
(lower part of Fig.~\ref{GAMMAJET:jetratio}). Although the width is still large for  
20~GeV/\emph{c} jets,  imposing a higher $p_{T}$ threshold produces a  loss of 
essential information about the jet (2~GeV/\emph{c} is already 10\% of the jet energy).
Consequently, the jet energy was calculated by requiring a $p_{T}$ threshold of 0.5~GeV/$c$
for $pp$ collisions and of 2~GeV/$c$ for Pb-Pb collisions. Nevertheless, 
to construct the jet fragmentation functions, as described in next sections, all detected particles
with $p_{T}^{th}=0.5$~GeV/$c$ inside the cone were taken into account.

\begin{figure}
\begin{center}\includegraphics[%
  width=6cm,
  keepaspectratio]{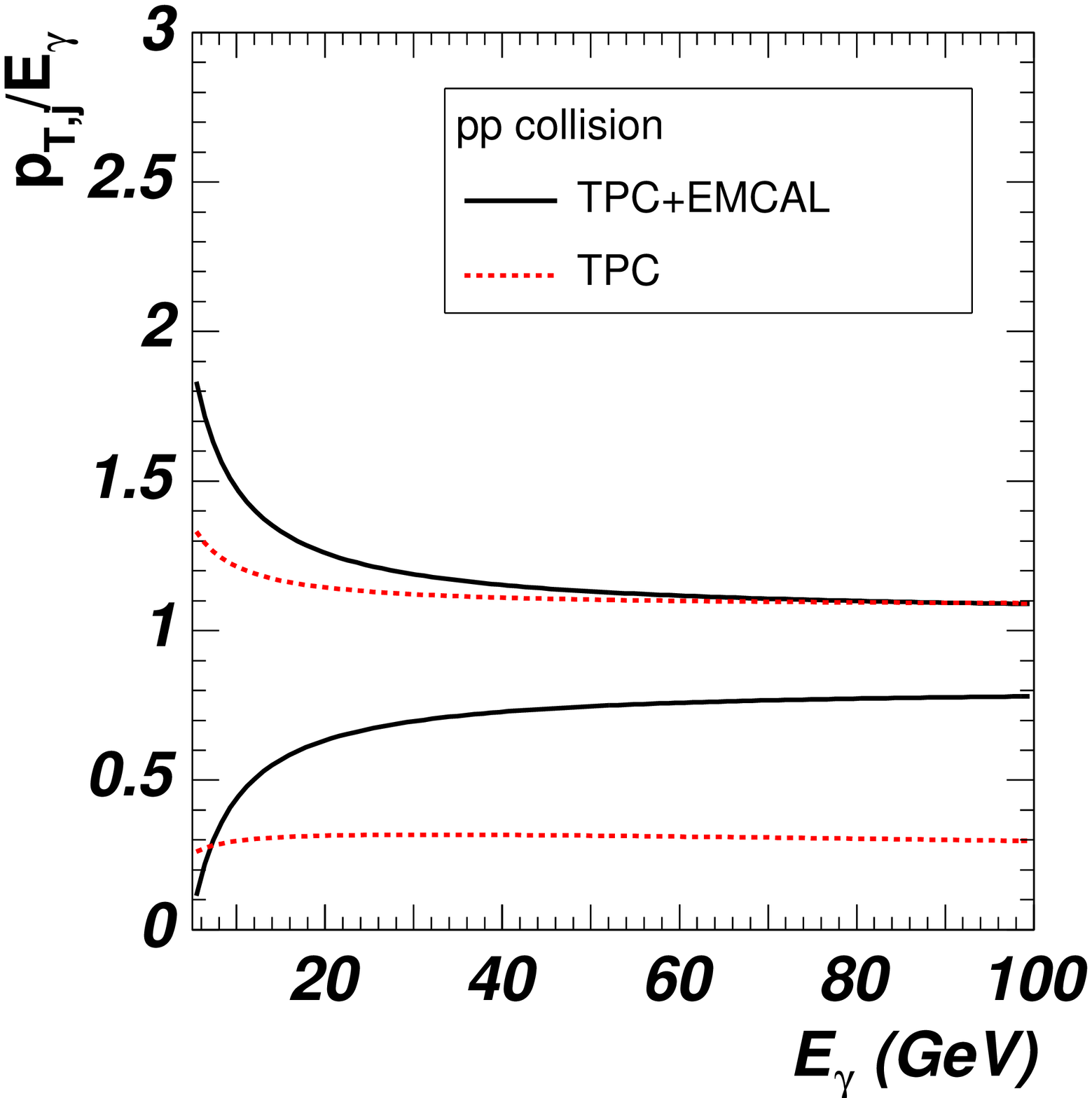}\end{center}

\begin{center}\includegraphics[%
  width=6cm,
  keepaspectratio]{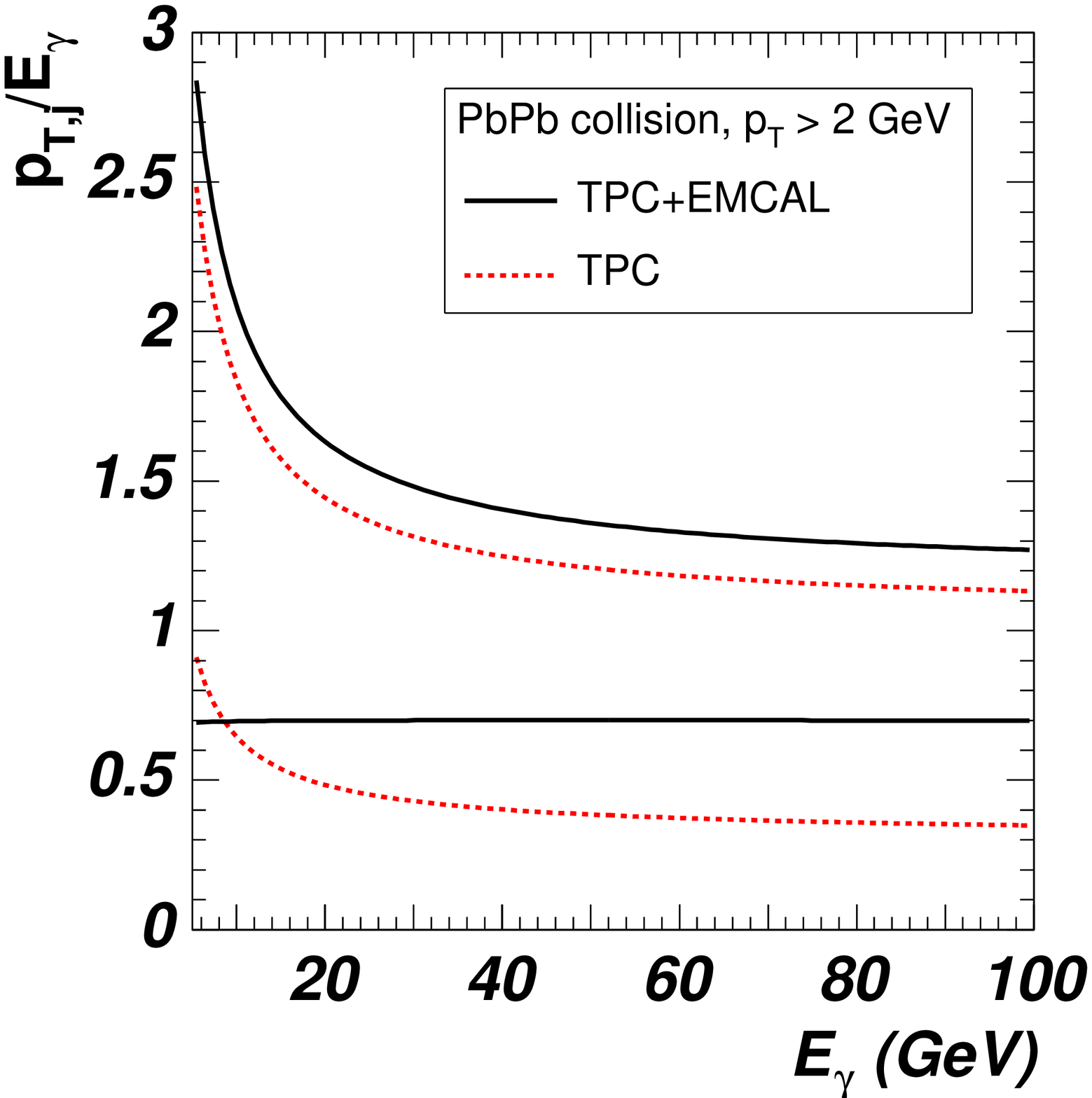}\end{center}

\caption{\label{GAMMAJET:jetratiolimit}{\small Maximum (upper line) and minimum
values (lower line) of the $p_{T,j}/E_{\gamma}$ ratio used to select
photon-jet events in the configurations with EMCal (solid line) and
without EMCal (dashed line) for $pp$ at $\sqrt{s}=5.5$~TeV (upper figure)
and Pb-Pb collisions at $\sqrt{s}=5.5A$~TeV (lower figure).}}
\end{figure}

Photon-jet events are well identified with the setup including EMCal 
when the ratio $p_{T,j}/E_{\gamma}$ is close to one. Two different values were considered
for the lower $p_{T,j}/E_{\gamma}$ limits, depending on the experimental setup (with or without
EMCal). The optimal selection windows (Fig.~\ref{GAMMAJET:jetratiolimit}), depend on 
the energy of the reconstructed jet.

The jet reconstruction algorithm fails for jets with $p_{T,j}<10$~GeV/$c$
due to the large fluctuations  of the  $p_{T,j}/E_{\gamma}$  ratio at low energies.
Therefore, such jets were excluded from our investigations.

\section{ Efficiency and contamination of jet selection\label{GAMMAJET:EfficCont} }

\begin{figure}
\begin{center}\includegraphics[%
  width=6cm,
  keepaspectratio]{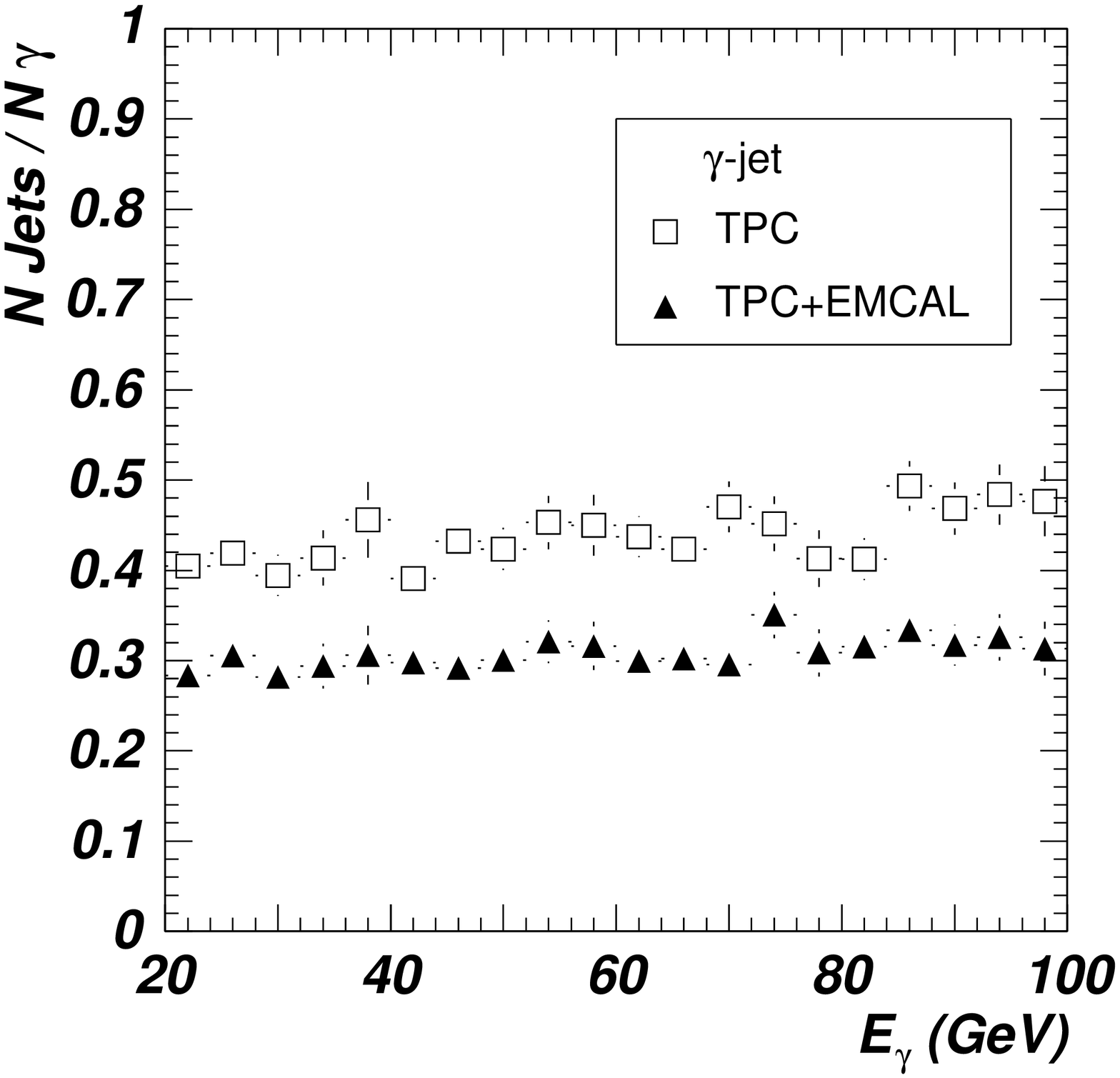}\end{center}

\begin{center}\includegraphics[%
  width=6cm,
  keepaspectratio]{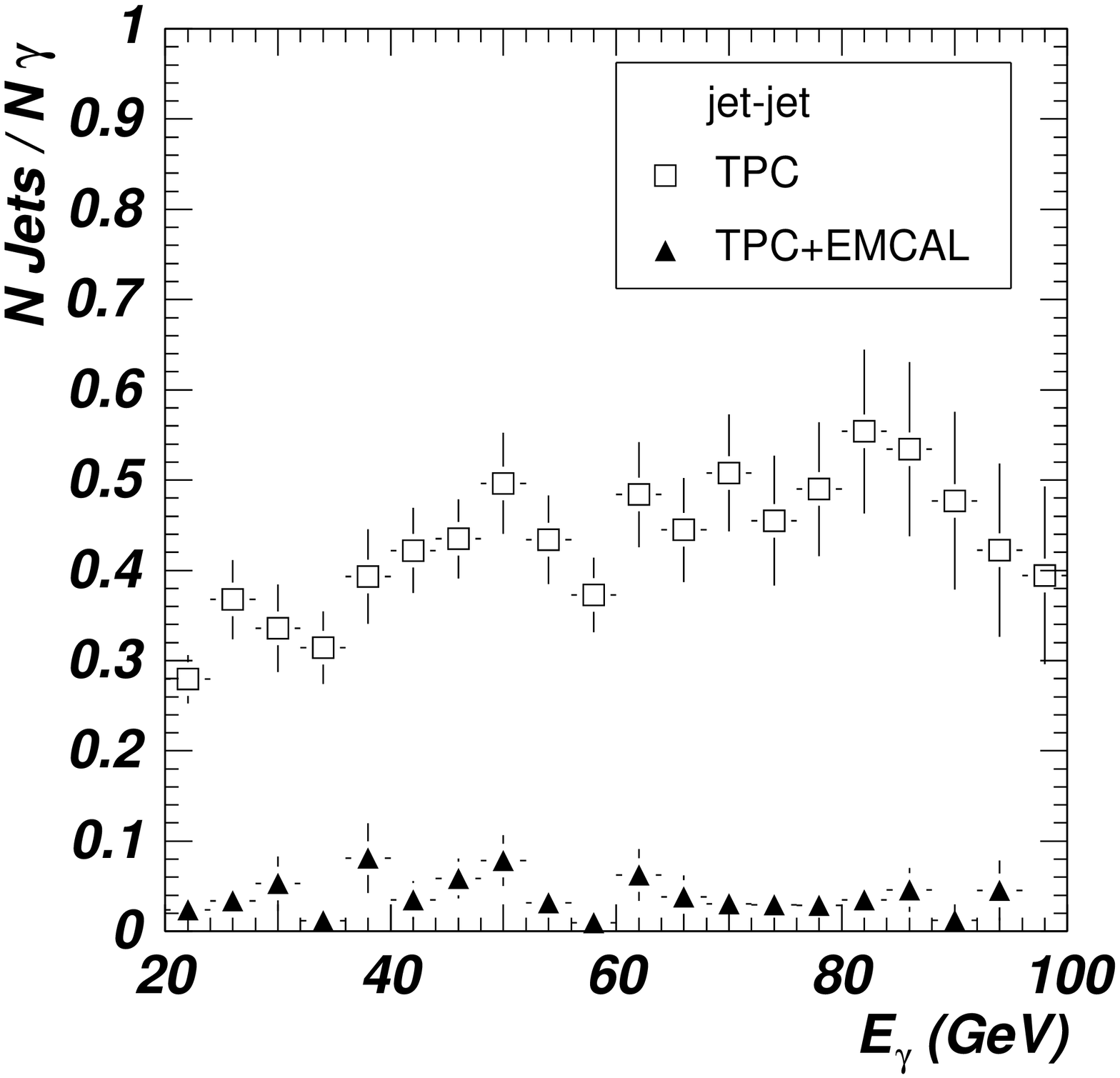}\end{center}

\caption{\label{GAMMAJET:effipp}{\small Upper frame: Jet selection efficiency.
Lower frame: Number of accepted jet-jet events divided by the number
of $\gamma$-like particles detected in PHOS (not identified). Results
for} \emph{\small pp} {\small collisions and for the setups without
EMCal~($\square$) and with EMCal~($\blacktriangle$).}}
\end{figure}

\begin{figure}
\begin{center}\includegraphics[%
  width=6cm,
  keepaspectratio]{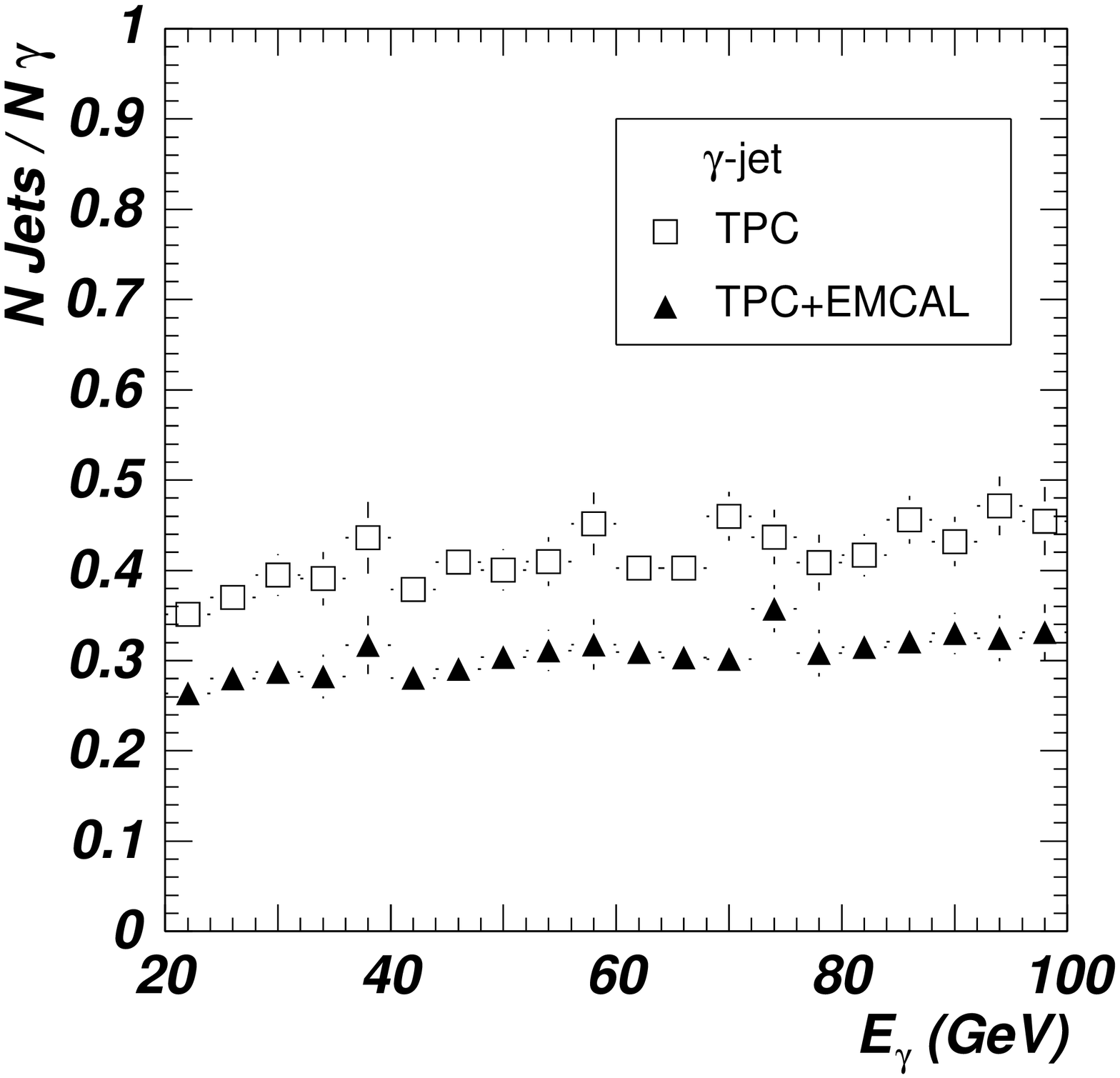}\end{center}

\begin{center}\includegraphics[%
  width=6cm,
  keepaspectratio]{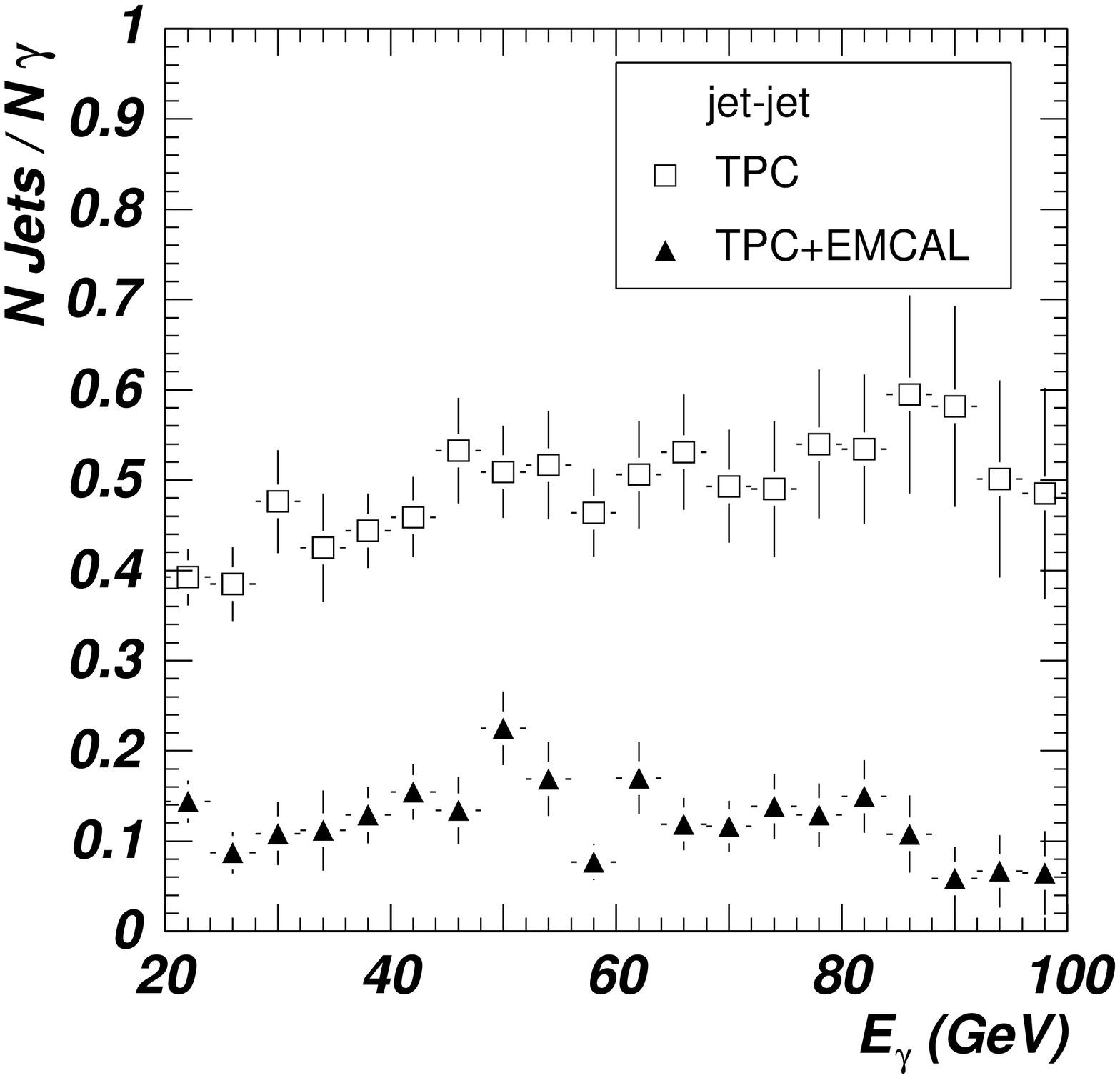}\end{center}

\caption{\label{GAMMAJET:effiPb}{\small Upper frame: Jet selection efficiency.
Lower frame: Number of accepted jet-jet events divided by the number
of $\gamma$-like particles detected in PHOS (not identified). Results
for Pb-Pb collisions at $\sqrt{s}=5.5A$~TeV and for the setups without EMCal~($\square$)
and with EMCal~($\blacktriangle$).}}
\end{figure}

The jet selection efficiency is defined as the ratio of the number
of identified $\gamma$-tagged jets to the number of prompt photons
detected in PHOS (Fig.~\ref{GAMMAJET:effipp} for $pp$ collisions 
and Fig.~\ref{GAMMAJET:effiPb} for Pb-Pb collisions).
The efficiency for the configuration with EMCal is about 30\%. For
the configuration without EMCal, the efficiency is higher (about 40-50\%)
due to the larger acceptance of the central tracking system. 
To estimate the contamination level of wrongly identified $\gamma$-jets 
from  jet-jet events, we  applied the $\gamma$-jet algorithm also to these events. 
This contamination  originates mainly from the decay photons contained in 
jets reaching PHOS, which may be misidentified as prompt photons and provide 
a seed for the algorithm. Indeed, the contribution of these background 
prompt photons can be even larger than the contribution of 
true prompt photons~\cite{ICM}. When the  $\gamma$-tagging algorithm is  
applied to jet-jet events, a substantial fraction of misidentified prompt photons 
is rejected. When no prompt photon identification is performed in PHOS, 
only about 10\% of the jet-jet events are accepted in the setup with EMCal but the 
value raises to 40-50\% in the absence of EMCal for both $pp$ and Pb-Pb collisions 
(Figs.~\ref{GAMMAJET:effipp} and \ref{GAMMAJET:effiPb}).

\begin{figure}
\begin{center}\emph{pp} collisions\end{center}

\begin{center}\includegraphics[%
  width=6cm,
  keepaspectratio]{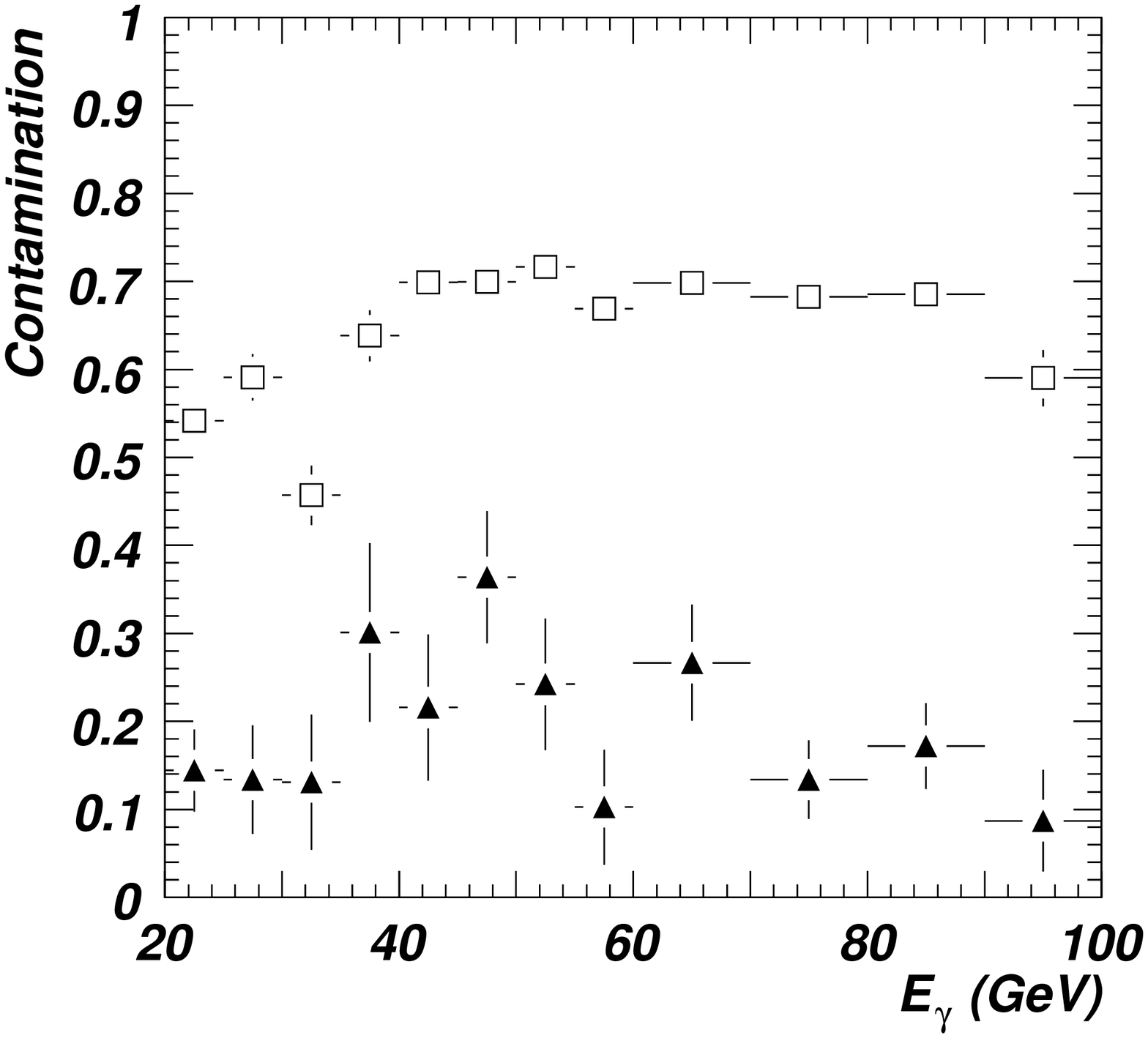}\end{center}

\begin{center}Pb-Pb collisions\end{center}

\begin{center}\includegraphics[%
  width=6cm,
  keepaspectratio]{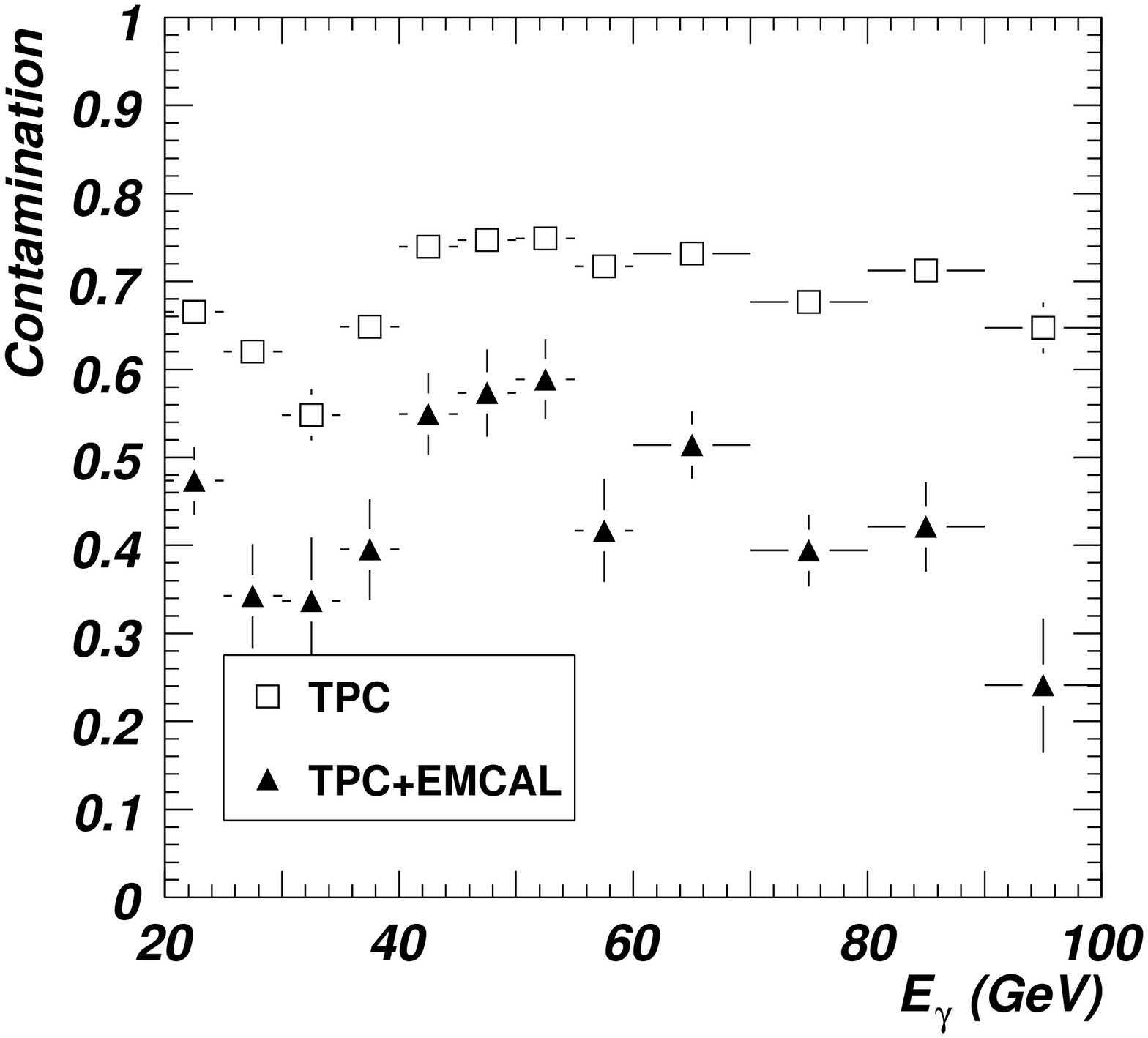}\end{center}

\caption{\label{GAMMAJET:purcont}{\small $\gamma$-tagging contamination
without prompt photon identification in PHOS, i.e., a high-$p_{T}$
neutral signal triggers the jet finding algorithm. Upper (lower) frame
corresponds to $pp$  (Pb-Pb) collisions at $\sqrt{s}=5.5A$~TeV. Results
shown for the setups without EMCal~($\square$) and with EMCal~($\blacktriangle$).}}
\end{figure}

\begin{figure}
\begin{center}\emph{pp} collisions\end{center}

\begin{center}\includegraphics[%
  width=6cm,
  keepaspectratio]{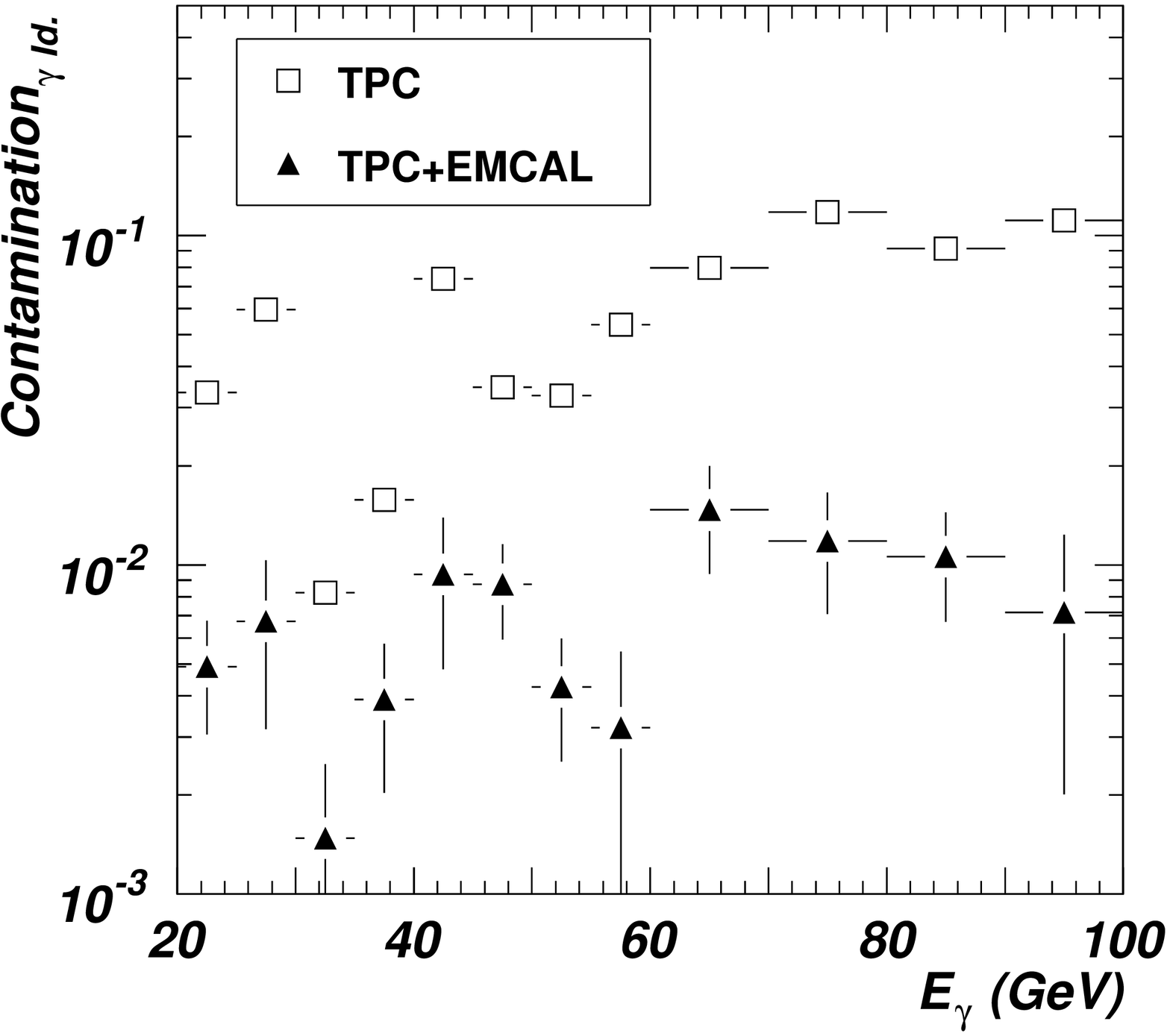}\end{center}

\begin{center}Pb-Pb collisions\end{center}

\begin{center}\includegraphics[%
  width=6cm,
  keepaspectratio]{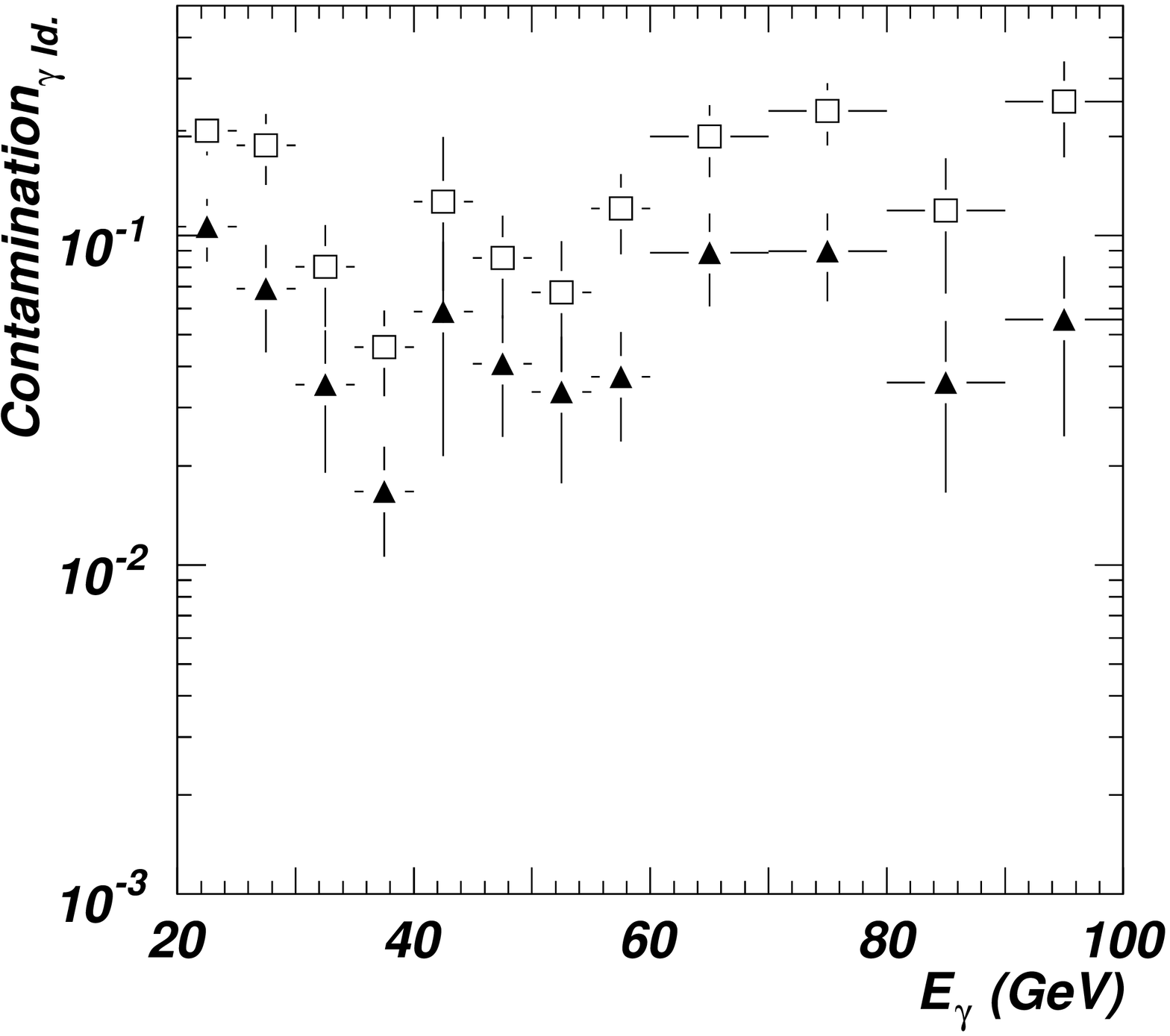}\end{center}

\caption{\label{GAMMAJET:idpurcont}{\small $\gamma$-tagging contamination
with prompt photon identification in PHOS (SSA+ICM, see definitions in  
Refs.~\cite{ICM,PPRv2}). Upper (lower) frame corresponds to $pp$  (Pb-Pb) collisions at 
$\sqrt{s}=5.5A$~TeV. Results shown for the setups without EMCal~($\square$) 
and with EMCal~($\blacktriangle$).}}
\end{figure}

We  studied the purity $\mathcal{P}$ and the contamination $\mathcal{C}$ 
of  $\gamma$-jet identified events. The purity is defined as the fraction of true 
$\gamma$-jet events among the identified events and the contamination as the fraction 
of jet-jet events identified as $\gamma$-jet events, i.e., $\mathcal{C}=1-\mathcal{P}$. 
In a first step, the $\gamma$-tagging algorithm was triggered by every high-$p_{T}$ 
neutral signal detected in PHOS. In the case of the TPC+EMCal configuration, purities of
about 80\% and 60\%  for \emph{pp} and Pb-Pb collisions, respectively, 
were obtained (Fig.~\ref{GAMMAJET:purcont}). Without EMCal, the purity was only of 20-40\%.  
In a second step, the prompt photon identification  was  switched on by including a 
shower shape analysis (SSA) and an isolation cut method (ICM)~\cite{ICM}. A strong 
enhancement of the purity was obtained (Fig.~\ref{GAMMAJET:idpurcont}). In the case of $pp$ collisions, the purity was larger than 90\% for the configuration without EMCal
and about 99\% for the configuration with EMCal. In the case of Pb-Pb collisions and 
for the configuration without EMCal  the purity levels were about 80\%  in
 the whole  energy range except  between 30 and 50 GeV where the purity was about 
 90\%. With EMCal the purity was higher than 90\% in the entire energy range.


\section{Fragmentation functions\label{GAMMAJET:correlation}}

\begin{figure}
\begin{center}\emph{pp} collisions\end{center}

\begin{center}\includegraphics[%
  width=6cm,
  keepaspectratio]{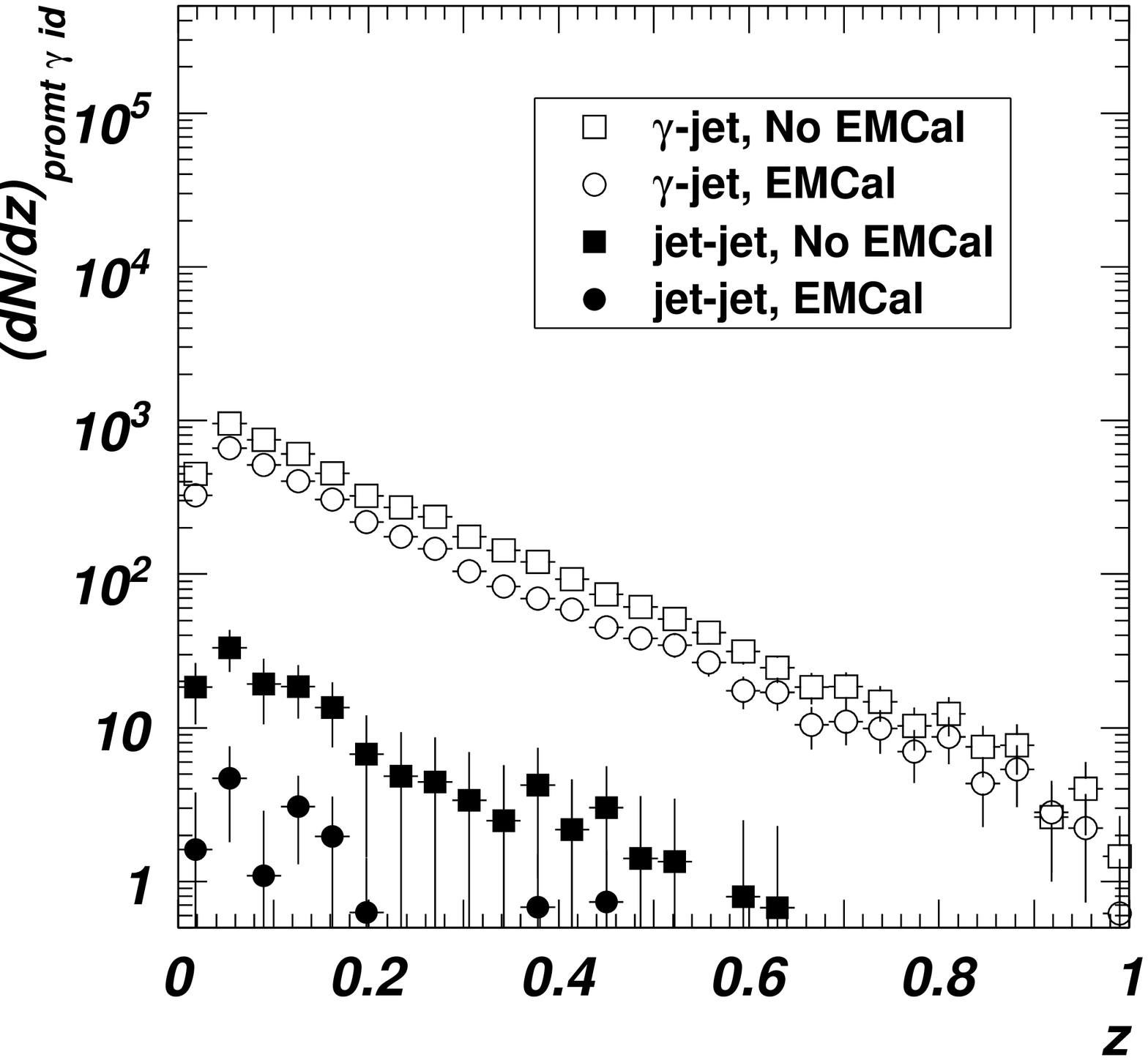}\end{center}

\begin{center}Pb-Pb collisions\end{center}

\begin{center}\includegraphics[%
  width=6cm,
  keepaspectratio]{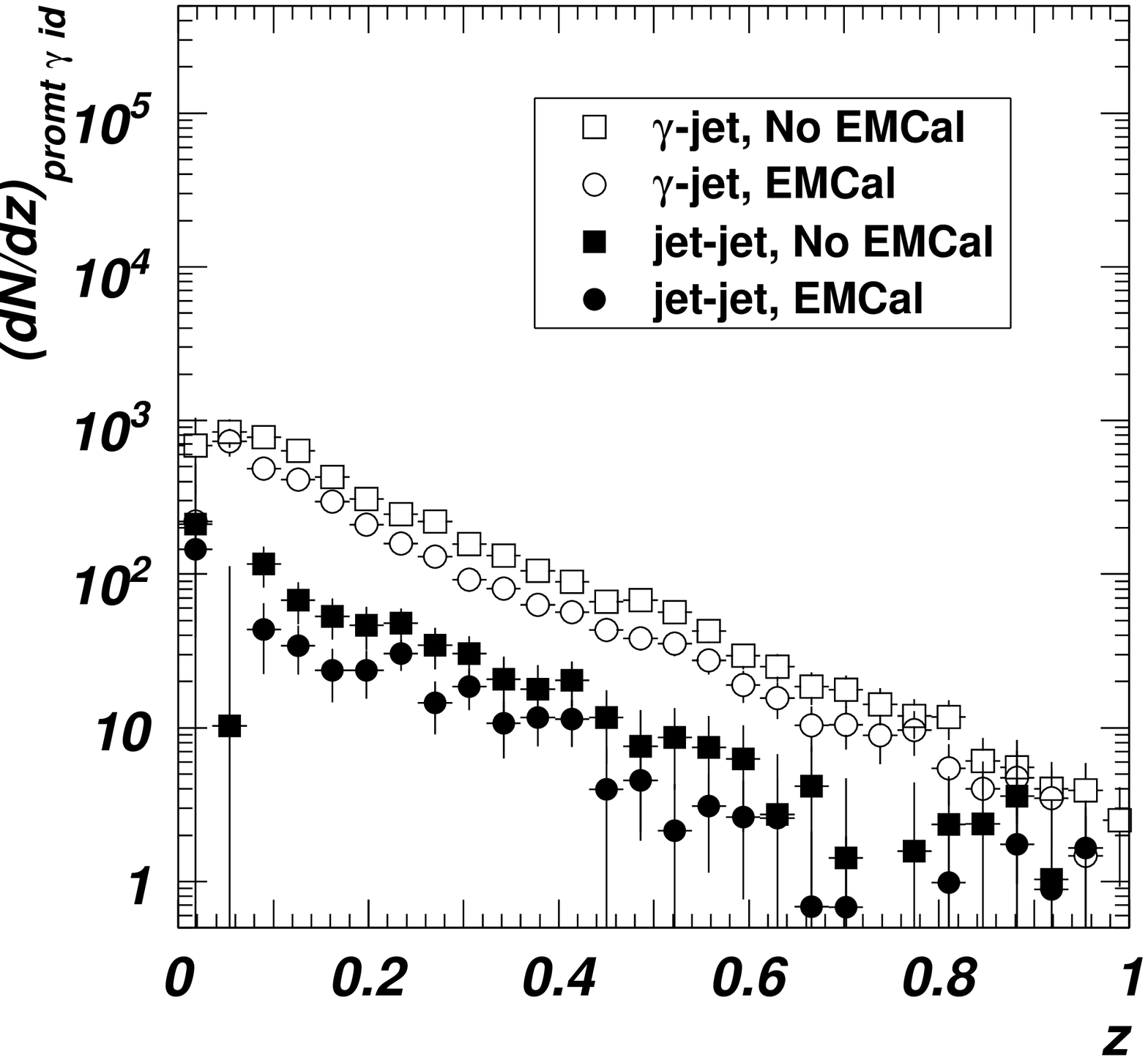}\end{center}

\caption{\label{GAMMAJET:ffjetpidic}{\small Fragmentation function for $\gamma$-jet
and jet-jet events with photon energy larger than 20~GeV for $ pp$ (upper frame) and Pb-Pb 
(lower frame) collisions at $\sqrt{s}=5.5A$~TeV. Prompt photons are identified in  PHOS by 
medium purity SSA and ICM ($pp$ collisions) or ICMS (Pb-Pb collisions),  
and  the heavy-ion background  is statistically subtracted (see Refs.~\cite{ICM,PPRv2} for details 
on identification definitions). }}
\end{figure}

A well established  method to study quantitatively the interaction of jets with the  
medium is to investigate the redistribution of fragmentation hadrons in phase 
space~\cite{Salgado:2003rv}, i.e., to measure the jet fragmentation function. 
The experimental fragmentation function is the distribution of charged hadrons 
within jets as a function of the variable $z$, defined for hard processes with a $\gamma$-jet
pair in the final state as $z=p_{T}/E_{\gamma}$. The statistics for the fragmentation function 
measurement during a standard year of LHC running was estimated for both 
$pp$ and Pb-Pb collisions at $\sqrt{s}=5.5A$~TeV (Figs.~\ref{GAMMAJET:ffjet} and
 \ref{GAMMAJET:ffjetpidic}). These fragmentation functions were constructed from  
$\gamma$-jet events and misidentified jet-jet events by integrating all events with
 identified prompt photon energy larger than 20~GeV/$c$. The following conclusions are obtained:
\begin{figure}
\begin{center}\emph{pp} collisions\end{center}

\begin{center}\includegraphics[%
  width=6cm,
  keepaspectratio]{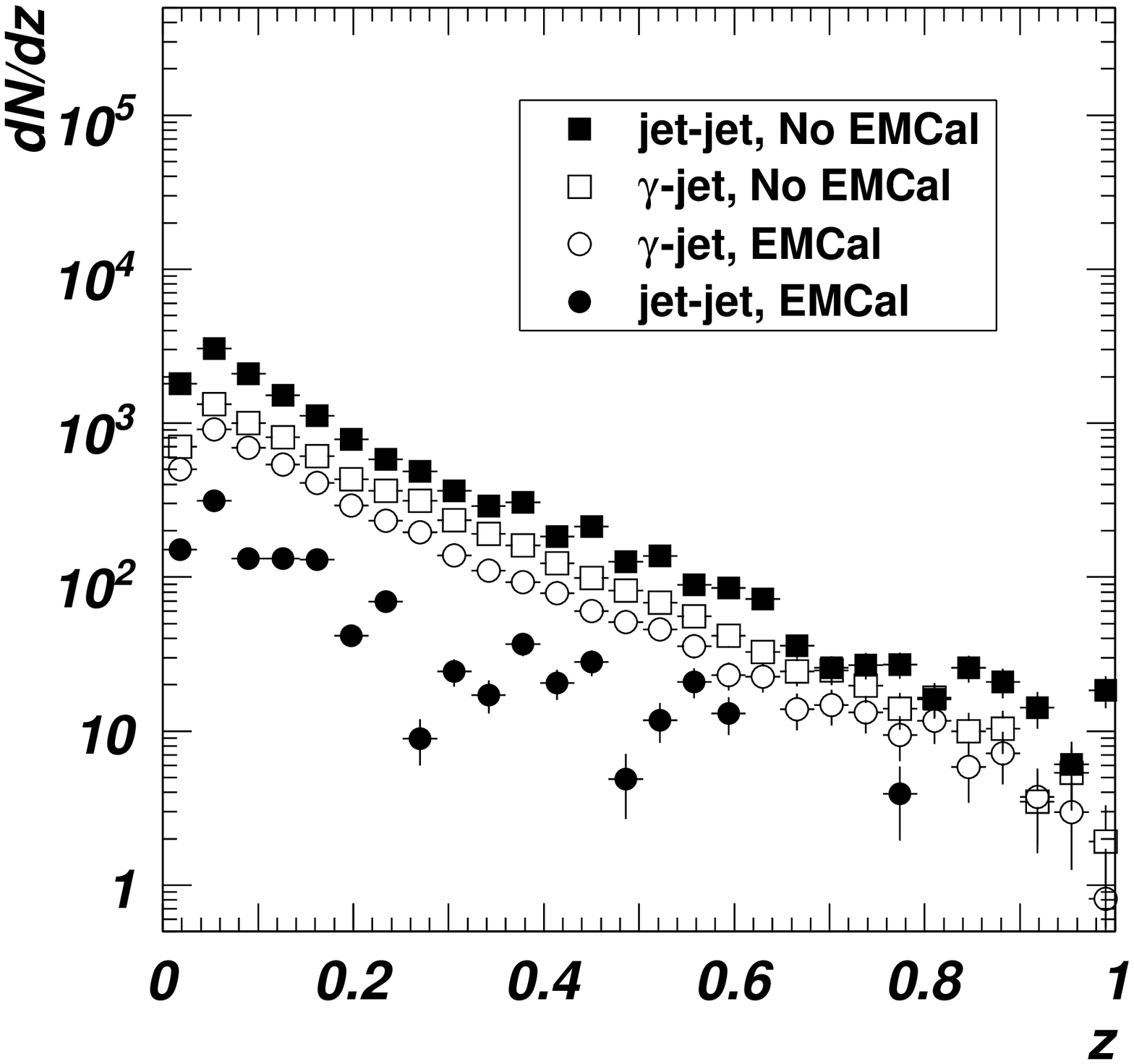}\end{center}

\begin{center}Pb-Pb collisions\end{center}

\begin{center}\includegraphics[%
  width=6cm,
  keepaspectratio]{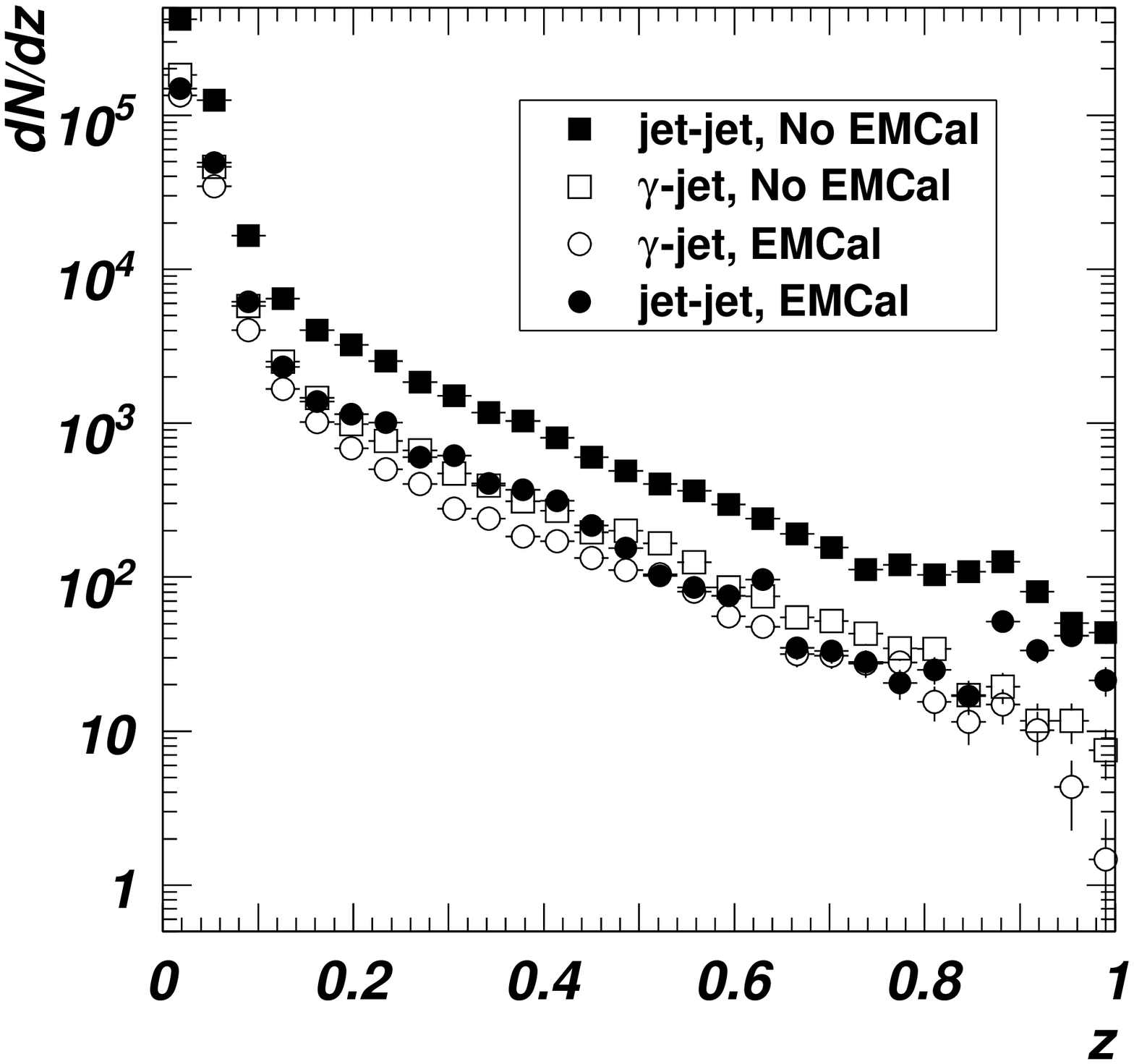}\end{center}

\caption{\label{GAMMAJET:ffjet}{\small Fragmentation function for $\gamma$-jet
and jet-jet events with photon energy larger than 20~GeV for $pp$ (upper frame) and 
Pb-Pb (lower frame) collisions at $\sqrt{s}=5.5A$~TeV. Neither prompt photon identification in 
PHOS nor heavy-ion environment background subtraction are done. }}
\end{figure}

\begin{enumerate}
\item In the case of \emph{pp} collisions, in the absence of EMCal and without  
prompt photon identification, the contamination due to misidentified jet-jet events 
dominates the true $\gamma$-jet contribution. In the setup with EMCal the contamination 
is suppressed, leading to a signal to background ratio close to 4. If prompt photons in PHOS 
are identified by medium purity SSA and ICMS (an isolation cut method with a threshold 
on the total $p_{T}$ sum)~\cite{ICM}, the contamination of misidentified jet-jet events 
is largely reduced, leading to a signal to background ratio of about 20 in the configuration 
without EMCal and near to 100\% background rejection for the setup with
EMCal. This reduction of background   is accompanied by   a reduction 
of the $\gamma$-jet statistics of only 10\%.
\item In the case of Pb-Pb collisions, the measured fragmentation function presents a peak 
at low \emph{z}, an artifact  due to low-$p_{T}$ charged hadrons coming from the 
heavy-ion underlying event. This peak can be statistically subtracted by calculating the 
contribution outside the cone defined by the leading particle. This contribution is estimated by
constructing a pseudo-fragmentation function with the particles inside a cone (with the same 
$R$ as the jet cone) centered at ($\phi_{\gamma}$, $\eta_{leading}$) which for $\gamma$-jet events is a region 
populated only by particles coming from the HIJING event with  no contribution from the jet. Prompt photon identification by medium purity SSA and ICM was required 
to reduce the contamination of wrongly identified jet-jet events to an acceptable level. The final 
signal to background ratio obtained is about 4 in the case without EMCal and rises up to 10 with 
EMCal.  The requirement of prompt photon identification reduces the statistics 
of $\gamma$-jet events by about 60\%.
\end{enumerate}

\begin{figure}
\begin{center}TPC+EMCal\end{center}

\begin{center}\emph{\includegraphics[%
  width=6cm,
  keepaspectratio]{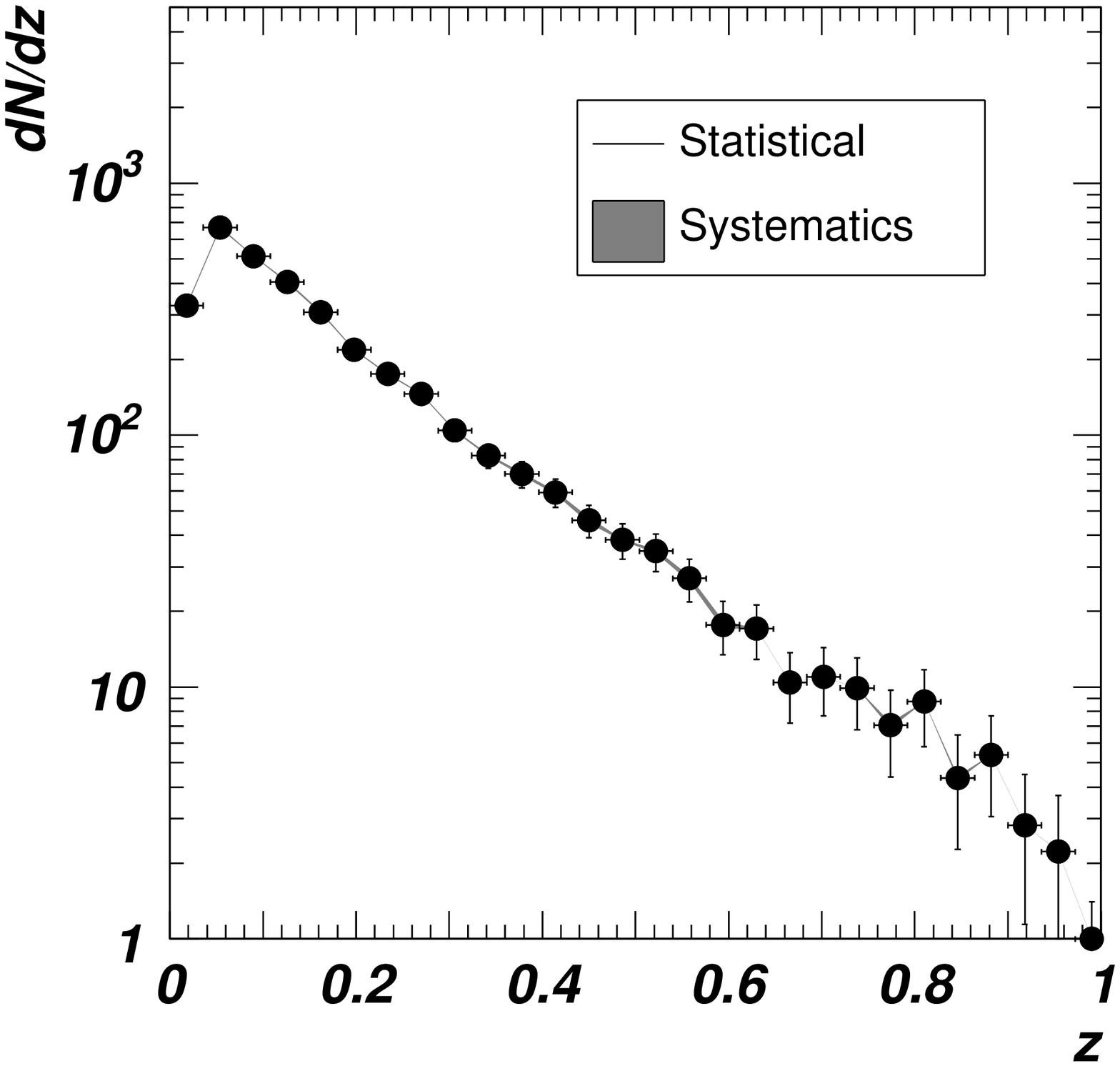}}\end{center}

\begin{center}TPC\end{center}

\begin{center}\emph{\includegraphics[%
  width=6cm,
  keepaspectratio]{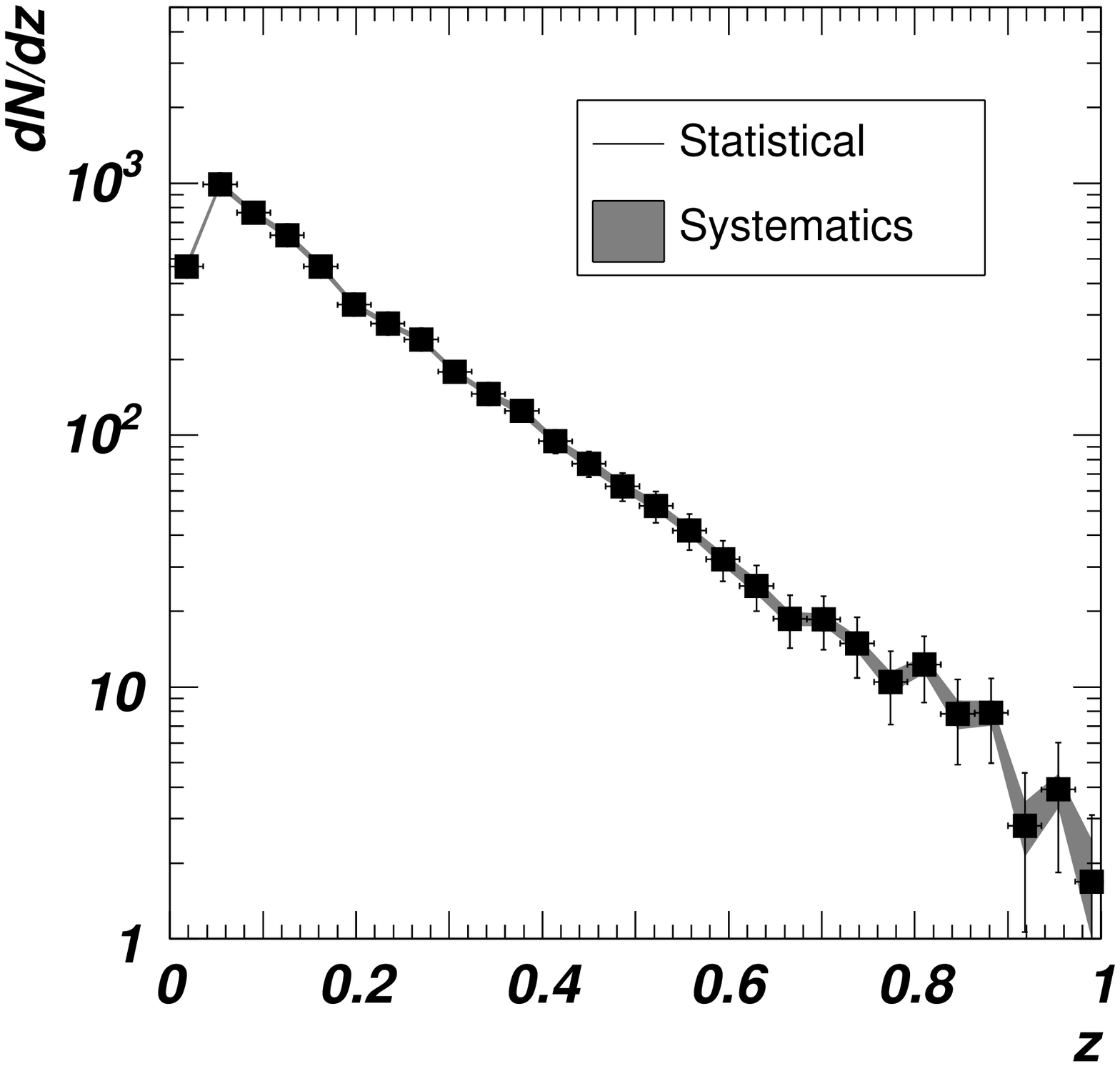}}\end{center}

\caption{\emph{\label{GAMMAJET:ffpp}}{\small Fragmentation function for 
$\gamma$-tagged jets ($\gamma$-jet + jet-jet events after prompt photon identification)
with photon energy larger than 20~GeV for a whole ALICE year, detected in
the central tracking system and EMCal (upper frame) and in the central
tracking system alone (lower frame), for $pp$ collisions. The shaded regions represent 
the systematic error due to jet-jet contamination. }}
\end{figure}

\begin{figure}
\begin{center}TPC+EMCal\end{center}

\begin{center}\emph{\includegraphics[%
  width=6cm,
  keepaspectratio]{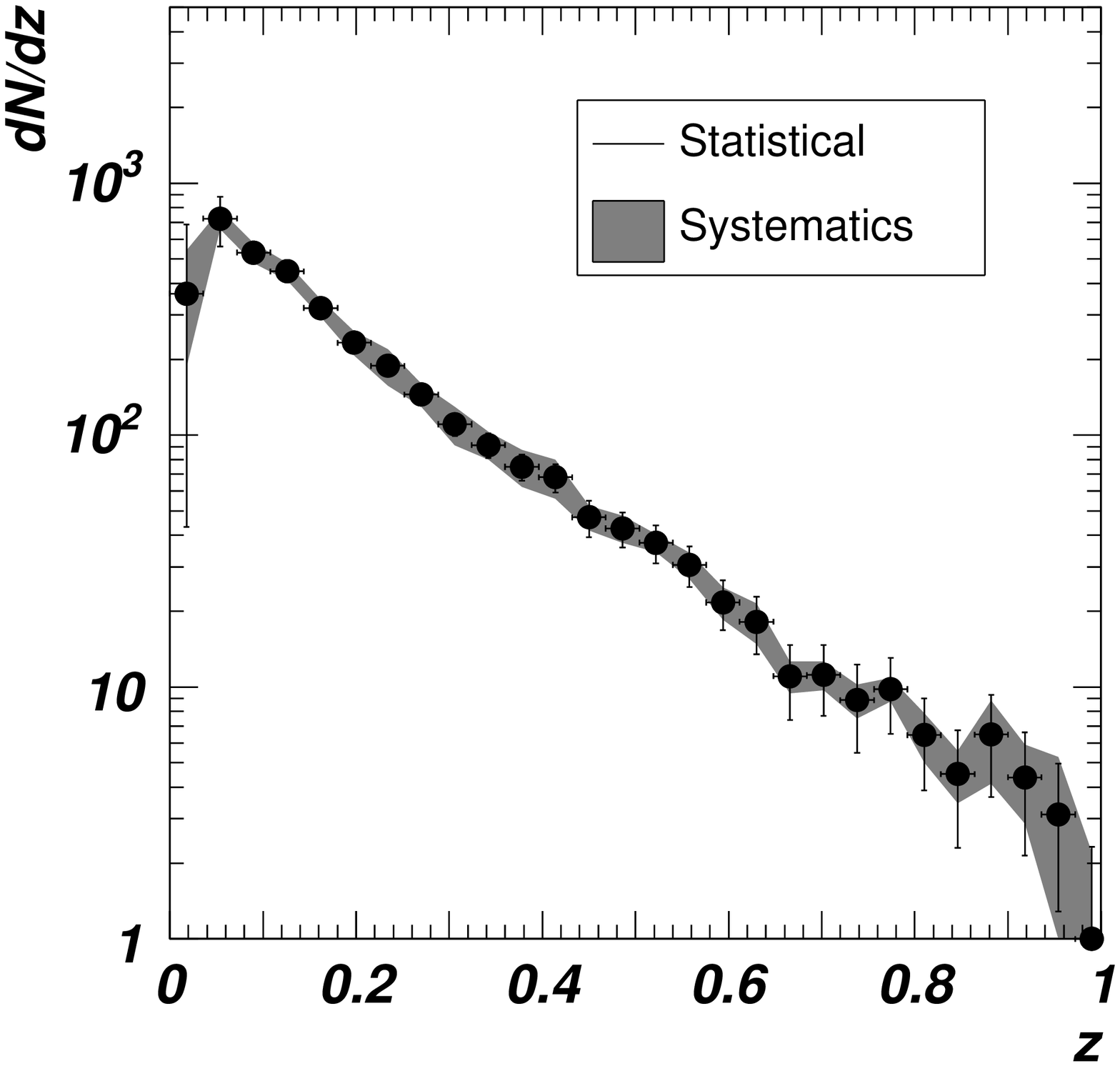}}\end{center}

\begin{center}TPC\end{center}

\begin{center}\emph{\includegraphics[%
  width=6cm,
  keepaspectratio]{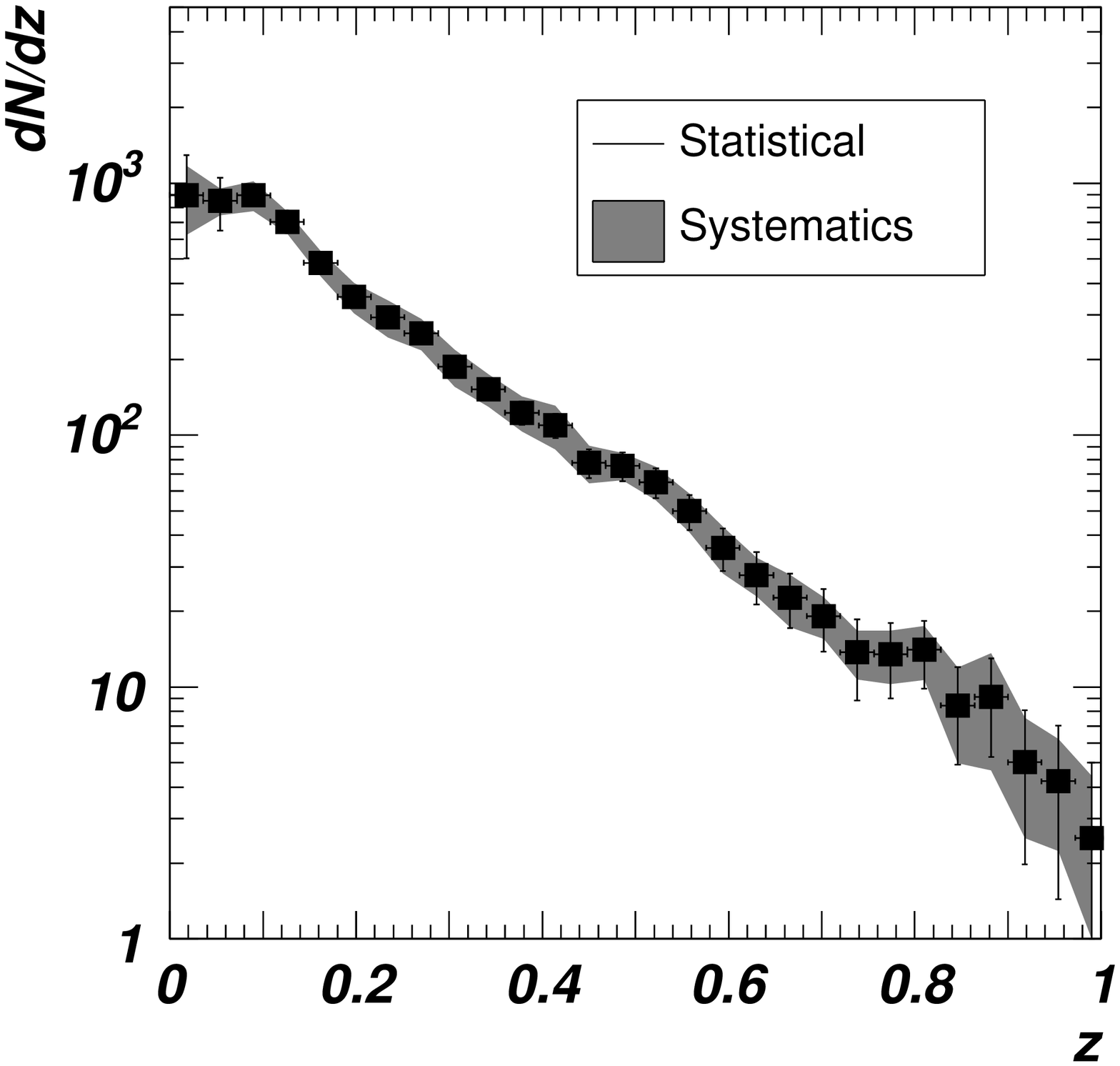}}\end{center}

\caption{\emph{\label{GAMMAJET:ffPb}}{\small Fragmentation function for $\gamma$-tagged
jets ($\gamma$-jet + jet-jet events after prompt photon identification)
with photon energy larger than 20~GeV for a whole ALICE year, detected in
the central tracking system and EMCal (upper frame) and in the central
tracking system alone (lower frame), for Pb-Pb collisions at $\sqrt{s}=5.5A$~TeV.
The shaded regions represent the systematic error due to jet-jet contamination. }}
\end{figure}
\begin{figure}
\begin{center}TPC+EMCal\end{center}

\begin{center}\emph{\includegraphics[%
  width=6cm,
  keepaspectratio]{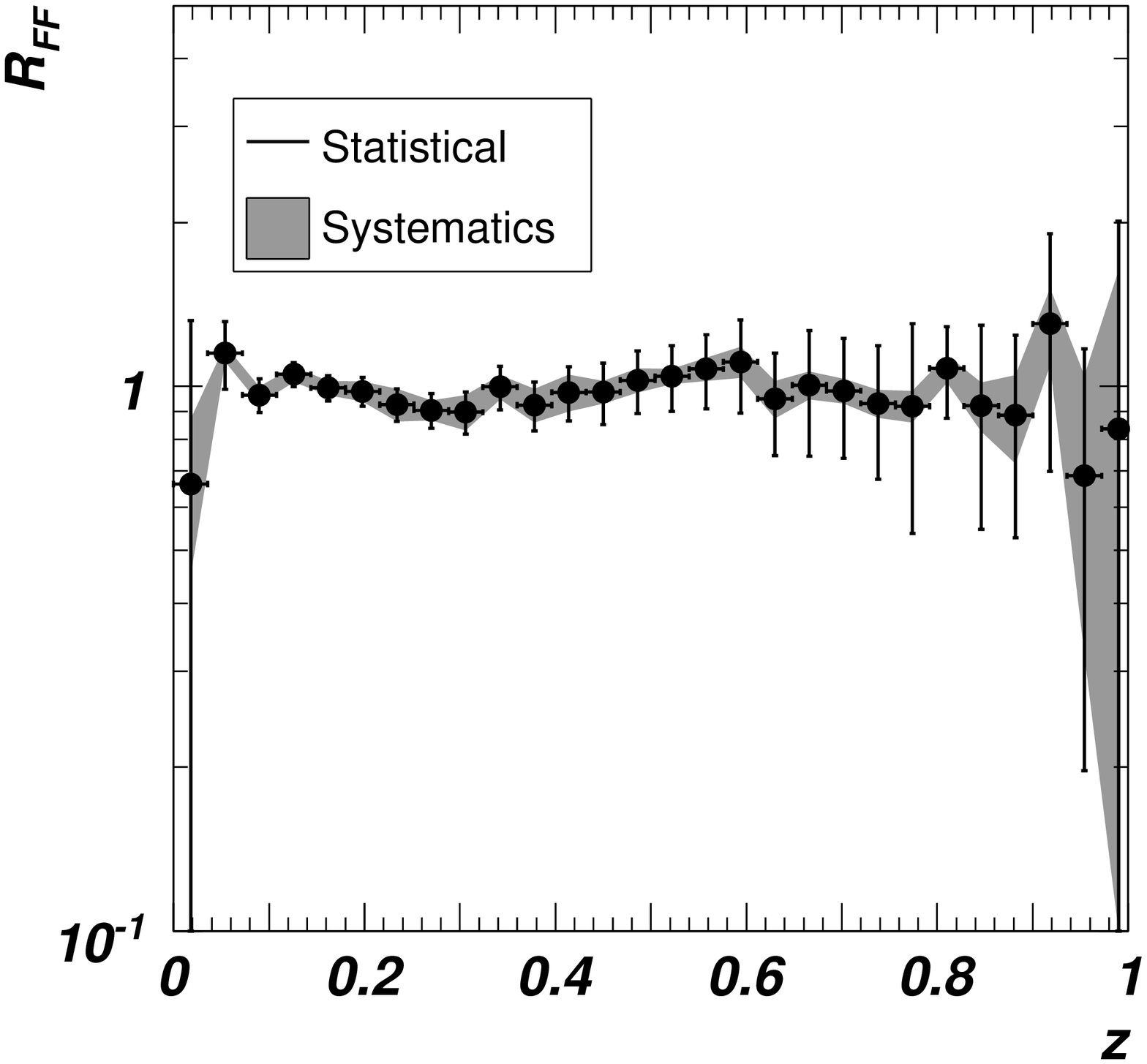}}\end{center}

\begin{center}TPC\end{center}

\begin{center}\emph{\includegraphics[%
  width=6cm,
  keepaspectratio]{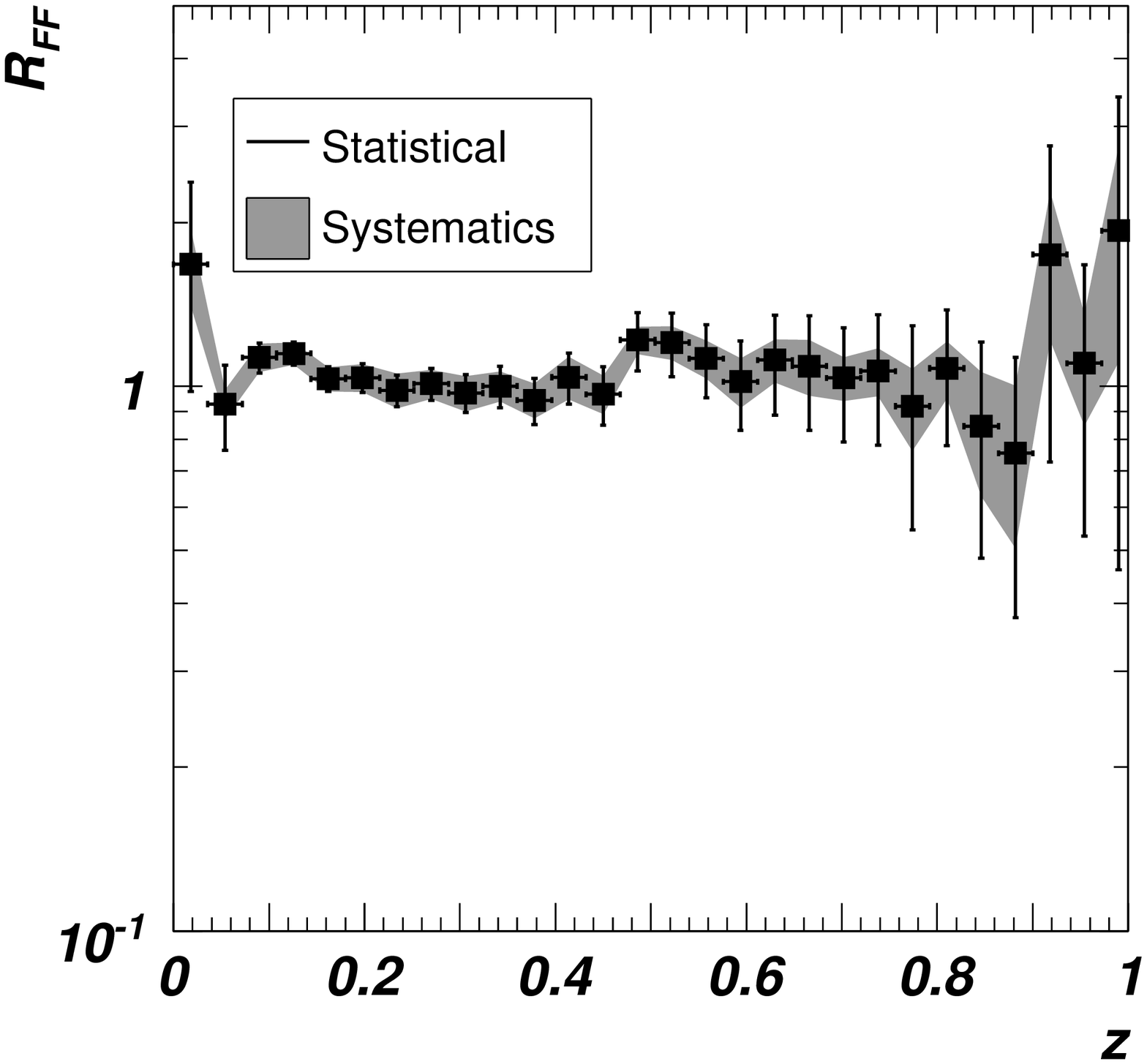}}\end{center}

\caption{{\small \label{GAMMAJET:ffPbpp}Ratio of the fragmentation functions
of $\gamma$-tagged jets with photon energy larger than 20~GeV for Pb-Pb
collisions at $\sqrt{s}=5.5A$~TeV scaled by the number of binary collisions~\cite{ICM} 
to $pp$ collisions at $\sqrt{s}=5.5$~TeV detected in the central tracking system and EMCal
(upper frame) or in the central tracking system alone (lower frame). The shaded region 
represents the systematic error due to  contamination from jet-jet events. }}
\end{figure}
\begin{figure}
\begin{center}TPC+EMCal\end{center}

\begin{center}\emph{\includegraphics[
  width=6cm,
  keepaspectratio]{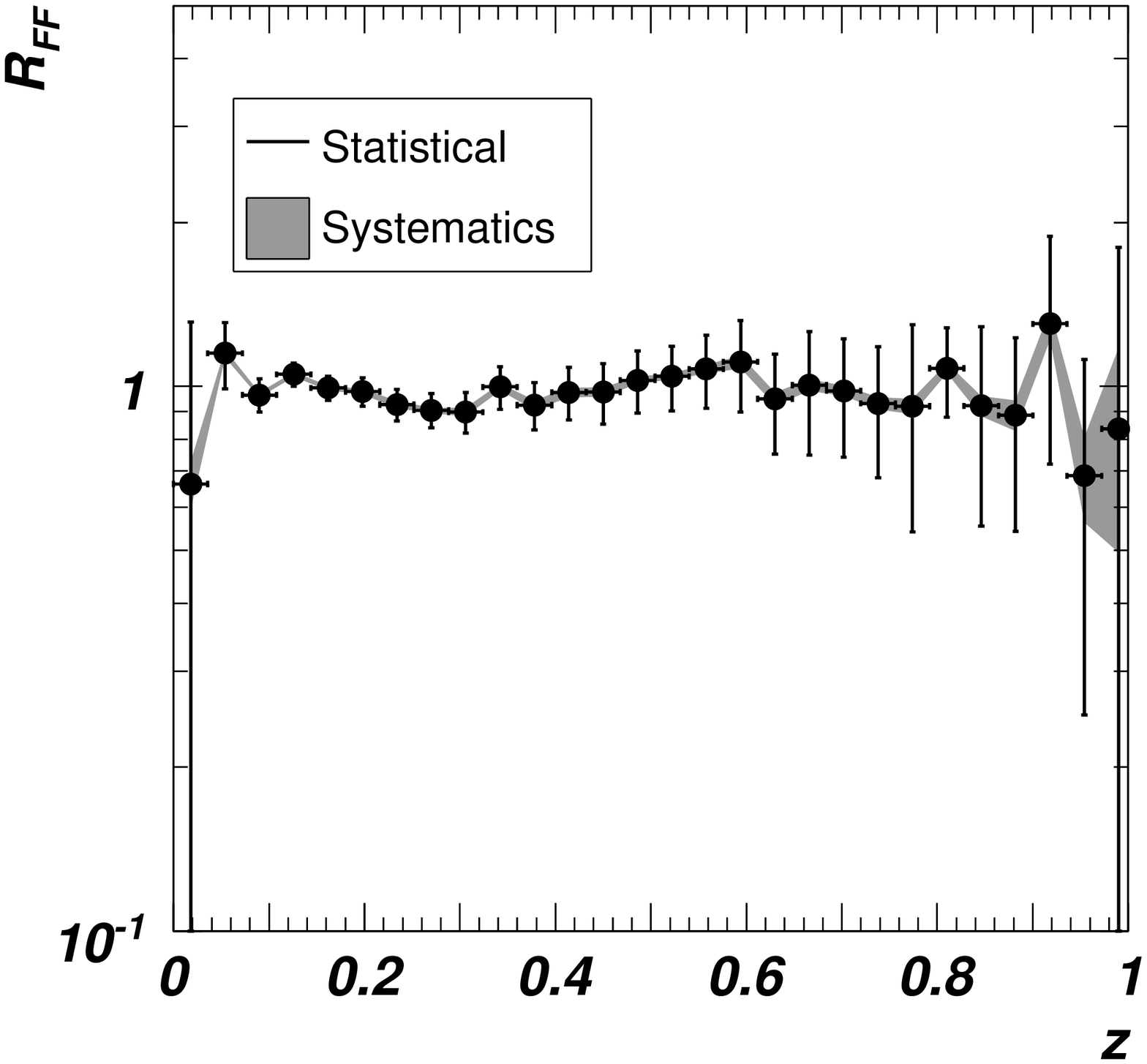}}\end{center}

\begin{center}TPC\end{center}

\begin{center}\emph{\includegraphics[
  width=6cm,
  keepaspectratio]{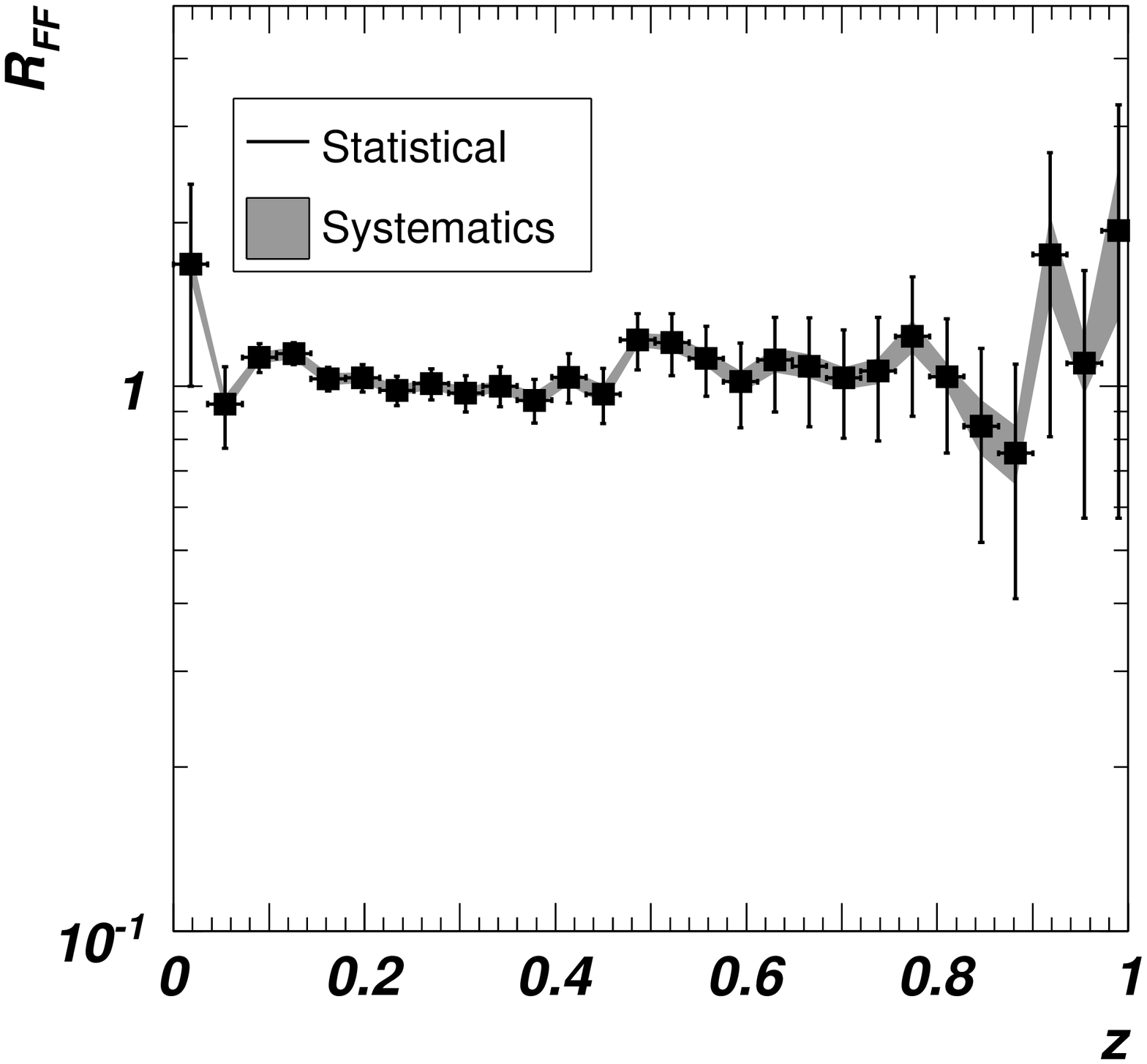}}\end{center}

\caption{{\small \label{GAMMAJET:ffPbppquench}Ratio of the fragmentation
functions of $\gamma$-tagged jets with photon energy larger than 20~GeV
for Pb-Pb collisions at $\sqrt{s}=5.5A$~TeV scaled by the number of binary 
collisions~\cite{ICM} to $pp$ collisions at $\sqrt{s}=5.5$~TeV detected in the central tracking 
system and EMCal (upper frame) or in the central tracking system alone (lower frame).
The shaded region represents the systematic error due to the contamination
from jet-jet events. Background is quenched a factor 5 like at RHIC.}}
\end{figure}

The fragmentation functions were calculated as the sum of the
identified $\gamma$-jet events and jet-jet background events scaled
by the  corresponding cross section (Figs.~\ref{GAMMAJET:ffpp} and
\ref{GAMMAJET:ffPb}). Statistical errors were calculated from the statistics that can be 
accumulated in a standard LHC year of running. Systematic errors reflect the amount 
of contamination. For Pb-Pb collision, in the range $0.1<z<0.5$, systematic and 
  statistical errors are of  the same order,  both with and without EMCal.  Statistical errors 
outside this \emph{z} range are too large to measure medium modifications effects. 
The inclusion of EMCal reduces the systematic errors by a factor of about 5 in $pp$ 
collisions and about 2 in Pb-Pb collisions. When  high $p_T$ $\pi^0$ suppression as 
observed at RHIC~\cite{Adams:2005dq, Tonjes:2004nz, Arsene:2004fa, Adcox:2004mh} 
is accounted for,  systematic errors are additionally reduced by a factor 5, becoming thus
systematically smaller than the statistical errors,  both  with and without EMCal. 

The sensitivity of photon-tagged jet fragmentation functions to nuclear medium modifications
can be estimated from the nuclear modification factor $R_{FF}$. This factor  is defined as 
the ratio of the fragmentation functions measured in \emph{AA}  and  \emph{pp} collisions 
scaled to the number of binary \emph{NN} collisions, with both fragmentation functions 
calculated for the same beam luminosity and running time. $R_{FF}$  should be equal to unity in
the absence of nuclear effects. As no medium modification effect was included in our simulations,
we obtain indeed  a value close to one over the entire \emph{z} range 
(Fig.~\ref{GAMMAJET:ffPbpp}). The statistical and systematic errors indicate that variations 
of $R_{FF}$ larger than 5\% could be measured for both setups in the range $0.1<z<0.5$. 
If $\pi^0$ are quenched by a factor of 5, as observed at RHIC, the systematic error would be
smaller than 5\% for both setups, as displayed in Fig.~\ref{GAMMAJET:ffPbppquench}. 
However,  there is no possibility to  measure the nuclear modification factor with an accuracy
better than 5\% due to limited $\gamma$-jet  statistics expected in one year of data taking.

We  may still consider another measurement approach in which prompt photons are detected 
by EMCal and  jets are only measured by the central tracking system%
\footnote{ It is not advisable to use PHOS as a detector of jet neutral particles
due to its reduced acceptance.}. 
In this setup, the prompt photon counting rate would be enhanced by a factor of 7 and 
consequently the statistical errors would be reduced by a factor 2.6 (figures obtained 
by assuming  the responses of  PHOS and EMCal  identical). This reduction of statistical 
errors may allow to measure medium modification effects over a wider \emph{z} range. 

\section{Conclusions}
We have  developed an algorithm to identify $\gamma$-jet events generated
in \emph{pp} and Pb-Pb collisions at $\sqrt{s}=5.5A$~TeV with the LHC detector ALICE. 
The $\gamma$-jet events are identified by selecting a prompt photon in PHOS and searching
for the leading particle in the opposite direction inside the ALICE central tracking system. 
Two different setups, with and without the electromagnetic calorimeter EMCal, are 
considered. Jets are reconstructed by an algorithm in which the leading particle
is used as a seed. The reconstructed jet is correlated to the photon if a number of 
conditions are fulfilled. The efficiency of identifying $\gamma$-jet events is mainly 
determined by the acceptance of the central tracking system and EMCal. For jets of 
energy larger than 20~GeV, this efficiency is found to be from around 
40\% to 50\% with increasing photon energy without EMCal, and, due to the smaller EMCal
acceptance,  about 30\% with EMCal. A large contribution to the $\gamma$-jet
event background comes from   $\pi^{o}$ decay  photons misidentified in PHOS as 
prompt photons. With our $\gamma$-jet tagging method in combination with isolation cut analysis 
for  prompt photon identification, the misidentification of jet-jet events as $\gamma$-jet events 
in $pp$ collisions is of the order of  5\% without EMCal and less than 1\% with EMCal, and  in 
Pb-Pb collisions is of the order of  20\% without EMCal and around 10\% with EMCal. 
We  obtain from our
simulations that fragmentation functions could be  measured with sufficient accuracy
to obtain the nuclear modification factor $R_{FF}$ with  errors  low enough to probe 
medium modifications.  As a main conclusion, we 
find that nuclear medium modifications could be measured if they produce
variations of $R_{FF}$  larger than 5\% in the region $0.1<z<0.5$.

\section{Acknowledgment}
This work has been supported in part by the Spanish DGICYT under contract
 FPA2003-07581-C02-01.  One of us (G. C.) thanks Ministerio de Educaci\'on y Ciencia for a PhD fellowship contract FP200-5452 and the Marie Curie EU program for the ``Training Site'' contract HPMT-CT-2001-00346. Also, we thank  the support of the INTAS grant 03-52-5747.

\end{document}